\def\la{\mathrel{\hbox{\rlap{\hbox{\lower4pt\hbox{$\sim$}}}\hbox{$<$}}}}
\def\ga{\mathrel{\hbox{\rlap{\hbox{\lower4pt\hbox{$\sim$}}}\hbox{$>$}}}}

\def\kms{km~s$^{-1}$}
\def\vp{$v_{phot}$}
\def\tb{$T_{bb}$}
\def\lam{$\lambda$}

\documentclass[12pt,preprint]{aastex}

\shortauthors{Branch et al.}
\shorttitle{Type Ib Supernovae}

\begin{document}
\title {Direct Analysis of Spectra of Type Ib Supernovae}

\author {{David Branch\altaffilmark{1}}, {S.~Benetti\altaffilmark{2}}, {Dan
Kasen\altaffilmark{3}}, {E.~Baron\altaffilmark{1}},
{David~J. Jeffery\altaffilmark{4}}, {Kazuhito Hatano\altaffilmark{5}},
{R.~A. Stathakis\altaffilmark{6}},
{Alexei~V. Filippenko\altaffilmark{7}}, {Thomas
Matheson\altaffilmark{8}}, {A.~Pastorello\altaffilmark{9}},
{G.~Altavilla\altaffilmark{2,9}}, {E.~Cappellaro\altaffilmark{2}},
{L.~Rizzi\altaffilmark{2,9}}, {M.~Turatto\altaffilmark{2}}, {Weidong
Li\altaffilmark {7}}, {Douglas~C. Leonard\altaffilmark{10}}, and
{Joseph~C. Shields\altaffilmark{11}}}

\altaffiltext{1}{Department of Physics and Astronomy, University of
Oklahoma, Norman, Oklahoma 73019, USA}

\altaffiltext{2}{Osservatorio Astronomico di Padova, vicolo
dell'Osservatorio~5, I--35122 Padova, Italy}

\altaffiltext{3}{Department of Physics, University of California,
Berkeley, CA, 94720}

\altaffiltext{4}{Astrophysics Research Center, Department of Physics,
New Mexico Institute of Mining and Technology, Socorro,~NM 87801}

\altaffiltext{5}{Research Center for the Early Universe, Tokyo, Japan}

\altaffiltext{6}{Anglo--Australian Observatory, PO Box 296, Epping,
NSW 1710, Australia}

\altaffiltext{7}{Department of Astronomy, University of California,
Berkeley, CA 94720--3411}

\altaffiltext{8}{Harvard--Smithsonian Center for Astrophysics, 
60 Garden Street, Cambridge, MA 02138}

\altaffiltext{9}{Dipartimento di Astronomia dell'Universit\'a di
Padova, Vicolo dell'Osservatorio 2, 35122 Padova, Italy}

\altaffiltext{10}{Five College Astronomy Department, University of
Massachusetts, Amherst, MA 01003--9305}

\altaffiltext{11}{Physics and Astronomy Department, Ohio University,
Athens, OH 45701}

\begin{abstract}

Synthetic spectra generated with the parameterized supernova
synthetic--spectrum code SYNOW are compared to photospheric--phase
spectra of Type~Ib supernovae (SNe~Ib).  Although the synthetic
spectra are based on many simplifying approximations, including
spherical symmetry, they account well for the observed spectra.  Our
sample of SNe~Ib obeys a tight relation between the velocity at the
photosphere, as determined from the Fe II features, and the time
relative to that of maximum light.  From this we infer that the masses
and the kinetic energies of the events in this sample were similar.
After maximum light the minimum velocity at which the He~I features
form usually is higher than the velocity at the photosphere, but the
minimum velocity of the ejected helium is at least as low as
7000~\kms.  Previously unpublished spectra of SN~2000H reveal the
presence of hydrogen absorption features, and we conclude that
hydrogen lines also were present in SNe~1999di and 1954A. Hydrogen
appears to be present in SNe~Ib in general, although in most events it
becomes too weak to identify soon after maximum light.  The
hydrogen--line optical depths that we use to fit the spectra of
SNe~2000H, 1999di, and 1954A are not high, so only a mild reduction in
the hydrogen optical depths would be required to make these events
look like typical SNe~Ib.  Similarly, the He~I line optical depths are
not very high, so a moderate reduction would make SNe~Ib look like
SNe~Ic.
      
\end{abstract}

\keywords{radiative transfer -- supernovae: general -- supernovae:
individual (SN~1954A, SN~1983N, SN~1984L, SN~1991ar,
SN~1996N, SN~1997dc, SN~1998T, SN~1998dt, SN~1999di, SN~1999dn,
SN~2000H)}

\section{INTRODUCTION}

Supernovae of Type~II are those that have obvious hydrogen lines in
their optical spectra.  Type~IIb supernovae have obvious hydrogen
lines around the time when they reach their maximum brightness
(hereafter just ``maximum light'') but later the hydrogen lines become
weak or even disappear.  Type~Ib supernovae do not have obvious
hydrogen lines but they do develop conspicuous He~I lines after
maximum light. Neither hydrogen nor He~I lines are conspicuous in the
spectra of Type~Ic supernovae.  Most or all events of these four types
--- II, IIb, Ib, and Ic --- are thought to result from core collapse
in massive stars. (Type~Ia supernovae, whose spectra lack hydrogen and
have a strong absorption feature produced by Si~II
$\lambda\lambda6347,6371$, are thought to have a fundamentally
different origin, as thermonuclear disruptions of accreting or merging
white dwarfs.)  For a recent review of supernova spectral
classification, including its historical development and with
illustrations of spectra of each type, see Filippenko (1997).

In this paper we are concerned with optical photospheric--phase
spectra of Type~Ib supernovae (SNe~Ib).  Until recently, good
photospheric--phase spectra had been published for only two SNe~Ib:
SN~1983N (Richtler \& Sadler 1983; Harkness et~al. 1987) and SN~1984L
(Harkness et~al. 1987).  The situation has improved substantially now
that Matheson et~al. (2001) have published spectra of the SNe~Ib that
were observed at the Lick Observatory during the 1990s.  These newly
available spectra, together with some additional previously
unpublished spectra that are presented in this paper, motivated us to
carry out a comparative study of SN~Ib spectra.  (Matheson et~al. also
present spectra of SNe~IIb and Ic, the study of which we defer to
separate papers.)

Our method is to compare the observed SN~Ib spectra with synthetic
spectra that we generate with the fast, parameterized, supernova
spectrum--synthesis code, SYNOW.  We refer to this approach, in which
the goal is to extract some constraints on the ejected matter from the
observations in an empirical spirit, as ``direct'' analysis --- to
distinguish it from the process of making very detailed
non--local--thermodynamic--equilibrium (non--LTE) calculations of
synthetic spectra based on supernova hydrodynamical models (e.g.,
Baron et al. 1999).  Issues that we can explore by means of our direct
analysis include (1) line identifications; (2) the extent to which
synthetic spectra calculated on the basis of simple assumptions can or
cannot account for observed SN~Ib spectra; (3) the degree to which the
rather homogeneous appearance of SN~Ib spectra, pointed out by
Matheson et~al. (2001), reflects a genuine physical homogeneity; (4)
the velocities at which the He~I lines form, compared to the velocity
at the photosphere as determined by the Fe~II lines; (5) whether
hydrogen lines are present and, when they are, the velocities at which
they form.

Previous work on the interpretation of photospheric--phase spectra of
SNe~Ib is briefly summarized in \S2.  The observed spectra that were
selected for this project are discussed in \S3 and those that have
not been published previously are displayed.  The synthetic spectrum
calculations are described in \S4, and comparisons with observed
spectra are presented in \S5.  The results are summarized and
discussed in \S6.

\section{PREVIOUS WORK}

The classic early paper on the interpretation of the
photospheric--phase spectra of SNe~Ib was that of Harkness
et~al. (1987), who calculated local thermondynamic equilibrium (LTE)
synthetic spectra for parameterized model supernovae having power--law
density structures and homogeneous chemical compositions, and compared
them to observed spectra of SNe~1984L and 1983N.  Some observed
features that appear in spectra of all supernova types were readily
attributed to Ca~II and Fe~II lines. Most of the remaining conspicuous
features were convincingly attributed to lines of He~I, even though
large ad hoc overpopulations of the highly excited lower levels of the
He~I lines had to be invoked in order to account for their presence.
The explanation for the overpopulations was later shown to be
nonthermal excitation and ionization caused by the decay products of
radioactive $^{56}$Ni and $^{56}$Co (Lucy 1991; Swartz et~al. 1993).

A feature of special interest in the spectra of SNe~1984L and 1983N
was an absorption near 6300~\AA\ that could not be attributed to
Fe~II, Ca~II, or He~I lines.  [Harkness et~al. (1987) referred to this
as ``the 6300~\AA\ absorption'' and so will we, although its
wavelength can be as short as 6200~\AA\ at early times.]  Harkness
et~al. tentatively attributed this absorption to C~II $\lambda6580$,
forming in outer high--velocity ($\ge$14,000~\kms) layers of the
ejected matter.  Later, on the basis of LTE synthetic spectra
calculated for radially stratified chemical compositions, Wheeler
et~al. (1994) suggested the absorption to be H$\alpha$, forming at
$\ge$13,000~\kms.  The issue of whether hydrogen is present in SNe~Ib
is very important because of its implications for the nature and
appearance of the progenitor stars, but it has been a difficult issue
to resolve because C~II
\lam6580, being less than 800~\kms\ to the red of H$\alpha$, is usually
a plausible alternative identification.

Woosley \& Eastman (1997) presented a comparison of a non--LTE synthetic
spectrum based on a particular explosion model with a
photospheric--phase spectrum of SN~1984L.  Overall, the synthetic
spectrum accounted rather well for the major features in the observed
spectrum.  The explosion model did not contain any hydrogen, and in
the synthetic spectrum the absorption nearest to the 6300~\AA\ feature
was produced by Si~II, but as we will see below this cannot be the actual
identification of the 6300~\AA\ absorption.  This is the only comparison
of an non--LTE synthetic spectrum with an observed SN~Ib
photospheric--phase spectrum to be published so far. 

No other SN~Ib was well observed until SN~1999dn.  Deng et~al. (2000)
used the same SYNOW code that we use in this paper to make a detailed
study of line identifications in three spectra that were obtained at
the Beijing Astronomical Observatory at times of about 10 days
before, at, and 14 days after maximum light.  In
addition to Fe II, Ca~II, and He~I lines, Deng et~al. explored the
possible role of lines of other ions (C~I, O~I, C~II, [O~II], Na~I,
Mg~II, Si~II, Ca~I, and Ni~II) in shaping the spectra of SN~1999dn.
They attributed the 6300~\AA\ absorption in the latest of their three
spectra of SN~1999dn to C~II $\lambda6580$, forming at
$\ge$10,000~\kms, but they suggested that in the earlier two spectra
the observed feature was more likely to be H$\alpha$, forming at
higher velocity.

\section{DATA}

The 11 SNe~Ib that were selected for this study are listed in Table~1.
An asterisk preceding the recession velocity, $cz$, indicates that it
is the value given by Matheson et al. (2001) for an H~II region near
the site of the supernova; otherwise the listed value is that of the
parent galaxy, from the Asiago Supernova Catalog (Barbon et~al. 1999;
updates are available at {\sl http://merlino.pd.astro.it/supern/}).
All observed spectra displayed in this paper are corrected for
redshift using the values of $cz$ listed in Table~1.  An asterisk
preceding the date of maximum light in the $V$ band, $t_{max}$,
indicates that only the date of discovery is listed, because the date
of maximum light is unknown.

Six of these events --- SNe~1991ar, 1997dc, 1998T, 1998dt, 1999di, and
1999dn --- were selected from Matheson et~al. (2001) because they
don't have obvious hydrogen lines while they do have conspicuous He~I
lines.  [SN~1991D, which also may be a Type~Ib but with fairly weak
He~I lines, is discussed in a separate paper (S.~Benetti et~al., in
preparation).] SNe~1998dt, 1999di, and 1999dn are especially useful for
our study because on the basis of photometry obtained at the Lick
Observatory, Matheson et al. were able to estimate the dates of
maximum light in the $R$ band, which we will assume to peak at the
same time as the $V$ band [as was the case for the Type~IIb SN~1996cb
(Qiu et~al. 1999)].  The three spectra of SN~1999dn that appeared in
Deng et~al. (2000) also are included in this study.

The spectra of SN~1983N are from Richter \& Sadler (1983) and Harkness
et~al. (1987), and the adopted date of maximum light, 1983 July~17,
(in the {\sl IUE} FES band, which is roughly like the $V$ band) is
from an unpublished manuscript that was circulated by N.~Panagia
et~al. in 1984.  The spectra of SN~1984L are from Harkness
et~al. (1987).  Tsvetkov (1987) estimated that SN~1984L reached
maximum light in the $B$ band on 1984 August~$20
\pm 4$ days. We assume that the $V$~band peaked two days later
(as was the case for SN~1996cb) and adopt August~22 as the date of
maximum light in the $V$ band.

SN~1954A is a special case because only photographic spectra, obtained
by N.~U.~Mayall at the Lick Observatory and by R.~Minkowski at the
Mount Wilson and Palomar Observatories, are available. Microphotometer
tracings of the spectra of SN~1954A and many other supernovae observed
at the Lick Observatory and the Mount Wilson and Palomar Observatories
between 1937 and 1971 have been digitized and displayed by Casebeer
et~al. (2000) and Blaylock et~al. (2000).  In the Asiago Catalog the
date of maximum light of SN~1954A in the $B$ band is estimated as 1954
April~19, so we will adopt April~21 for the $V$ band.

One previously unpublished spectrum of SN~1996N is included in this
study.  The spectrum, obtained at the Anglo--Australian Telescope on
1996 March~23 (Germany et~al. 2000), 11 days after discovery, is shown
in Figure~1.  The date of maximum light of SN~1996N is unknown.  The
spectrum appears to be that of a typical SN~Ib not long after maximum
light.

Six previously unpublished spectra\footnote{These spectra are
partially based on observations collected at the European Southern
Observatory, Chile, ESO~N$^0$65.H--0292, and at the Asiago
Observatory} of SN~2000H (Pastorello et~al. 2000; Benetti et~al. 2000)
also are included. These are shown in Figure~2.  The spectra of
SN~2000H resemble those of a typical SN~Ib except for an unusually
deep 6300~\AA\ absorption in the first four spectra, as well as a
weak, narrow absorption near 4650~\AA\ in at least the second and
third spectra. Benetti et~al. attributed these absorptions to
H$\alpha$ and H$\beta$.  (The H$\beta$ feature will be seen more
clearly in subsequent figures.)  This identification of hydrogen lines
might raise the question of whether SN~2000H should be regarded as a
Type~IIb, but we do not favor such a classification because even at
the earliest observed times the presence of hydrogen lines was not
obvious, as evidenced by initial classifications of SN~2000H as a
peculiar Type~Ia (Garnavich et~al. 2000) and a Type~Ic (Pastorello
et~al. 2000).  From unpublished ESO photometry of SN~2000H we estimate
that the date of maximum light in the $B$ band was 2000
February~$9\pm2$ days, so we adopt February~11 as the date of maximum
light in the $V$ band.

\section{CALCULATIONS}

Calculations have been carried out with the fast, parameterized
supernova spectrum--synthesis code, SYNOW.  Recent applications to
Type~Ic supernovae and brief descriptions of SYNOW can be found in
Millard et~al. (1999) and Branch (2001), and technical details of the
code are in Fisher (2000).  An extensive discussion and illustration
of the elements of supernova line formation appears in Jeffery \&
Branch (1990). The basic assumptions of SYNOW are spherical symmetry;
velocity proportional to radius; a sharp photosphere; and line
formation by resonant scattering, treated in the Sobolev
approximation.

Various fitting parameters are available.  The parameter $T_{bb}$ is
the temperature of the blackbody continuum from the photosphere.  The
values used in this paper range from 8500 to 3600~K and typically are
$\sim$6500~K around the time of maximum light and $\sim$5000~K
beginning roughly two weeks after maximum.  We do not attach much
physical significance to these values because (for one thing) the
observed spectra have not been corrected for interstellar extinction.

For each ion whose lines are introduced, the optical depth at the
photosphere of a ``reference line'' is a fitting parameter, and the
optical depths of the other lines of the ion are calculated assuming
Boltzmann excitation at excitation temperature $T_{exc}$.  In this
paper, to reduce the number of free parameters, we simply fix
$T_{exc}$ at a nominal SN~Ib value of 7000~K.  The relevant lines of a
given ion don't have widely differing excitation potentials so their
relative optical depths don't depend strongly on $T_{exc}$, within the
range of temperatures that are relevant here.

The line optical depths are taken to vary with ejection velocity as
$v^{-n}$.  Again for simplicity, in this paper we always use $n = 8$,
except for one illustration of the effects of using $n = 5$ instead.  In
the analysis of the SN~1999dn spectra by Deng et~al. (2000), the line
optical depths were taken to vary as $e^{-v/v_e}$, with
$v_e = 1000$~\kms.  Since the exponential distribution has an effective
power--law index of $n = v/v_e$, the distribution used by Deng
et~al. falls off more steeply than $n = 8$ for $v > 8000$~\kms\ and less
steeply for $v < 8000$~\kms.  This leads to some differences in the
values of $v_{phot}$ and the reference--line optical depths used by
Deng et~al. and by us to match the same observed spectra.

The maximum velocity of the line--forming region is set high enough so
that effectively there is no outer boundary. The default minimum
velocity of the line--forming region is the velocity at the
photosphere; when an ion is assigned a higher minimum velocity, that
ion is said to be detached from the photosphere.

Reasons that SYNOW spectra cannot be expected to provide exact fits to
observed spectra are numerous and obvious: the calculations are based
on many simplifiying assumptions, including spherical symmetry, and
the oscillator strengths (Kurucz 1993) are good but not perfect.  In
this paper we are not concerned with proposing a line identification
for every weak observed feature.  We are more interested in
establishing the identities of the major features and then
concentrating on a comparative analysis --- to investigate the degree
to which the SNe~Ib of our sample are physically similar, and to look
for differences.

\section{COMPARISONS}

\subsection{The Fiducial SN~Ib  Spectrum: SN1999dn, 17 Days After Maximum}

We begin the comparisons of observed and synthetic spectra by
concentrating on a ``fiducial'' SN~Ib spectrum --- a spectrum of a
typical SN~Ib that has good signal--to--noise ratio and broad
wavelength coverage, and in which most of the major spectral features
are well developed.  The best available spectrum for this purpose is
the Matheson et~al. (2001) spectrum of SN~1999dn obtained on
1999 September~17, 17 days after maximum light.  In
Figure~3, this spectrum is compared with a synthetic spectrum that has
\vp=6000~\kms\ and
\tb=4800~K, and contains lines of Fe~II, He~I, O~I, Ca~II, Ti~II, and
Sc~II.  Almost all of the features in the observed spectrum can be
attributed to these ions.  The discrepancies will be discussed as we
look at the contribution of each ion to the synthetic spectrum.  As
always when fitting observed spectra with SYNOW spectra, we are more
concerned with discrepancies in the wavelengths of absorption features
than with discrepancies in flux; the latter are inevitable given the
simplicity of our spectrum calculations.

Figure~4 is like Figure~3 but with nothing but the Fe~II lines in the
synthetic spectrum.  The optical depth of the reference line,
\lam5018, is 7.  The Fe~II lines are mainly responsible for the spectral
features from about 4300 to 5300~\AA, and they have additional effects
at shorter wavelengths. At this post--maximum time they may also be
responsible for the observed absorptions near 6100 and 6300~\AA.
(These two features are not strong enough in this particular synthetic
spectrum, but a higher value of $T_{exc}$ would increase their
strengths relative to the reference line.)  For this reason we don't
use H$\alpha$ or C~II \lam6580 to account for the weak 6300~\AA\
absorption in this observed spectrum.  Around maximum light, however,
Fe~II lines are not strong enough to account for the 6300~\AA\
absorption that is observed at that time.

In this paper we always determine the value of $v_{phot}$ on the basis
of the Fe~II features in the region from about 4300 to 5300~\AA.
Figure~5 shows a comparison of two Fe~II synthetic spectra that have
\vp = 5000 and 10,000~\kms. This figure shows that the spectral
signature of Fe~II in this wavelength range is quite sensitive to \vp.
Our fitting uncertainty in $v_{phot}$ is about 1000~\kms.

The top panel of Figure~6 is like Figure~3 but with only the He~I
lines in the synthetic spectrum.  The He~I lines are detached at
8000~\kms\ (recall that
\vp=6000~\kms), where the optical depth of the reference line,
\lam5876, is 10.  It is likely that two optical He~I lines, \lam 6678 and
\lam7065, are almost entirely responsible for their
corresponding observed features.  Two other lines, \lam5876 and
\lam4472, are mainly responsible for their corresponding observed
features but they may be blended with the Na~I D~lines
(\lam\lam5890,~5896) and Mg~II \lam4481, respectively.  In this
spectrum He~I \lam7281 accounts very nicely for an observed feature,
but in some other spectra the fit is not so good.  The remaining
optical He~I lines are weaker and in the synthetic spectrum of
Figure~3 they are overwhelmed by lines of other ions.  In Figures~3
and 6 the blue edge of the synthetic absorption produced by He~I
\lam10830 is not blue enough to account for the sharp drop in the
observed spectrum near 9000~\AA, but we show how to remedy this below.

In Figure~3 and the top panel of Figure~6 the He~I lines are detached
at 8000~\kms\ in order to fit the wavelengths of the corresponding
observed absorption features.  The detachment causes the flat tops of
the synthetic He~I emission components (which are superimposed on a
sloping continuum).  The rounded emission peak that is observed near
5900~\AA\ could easily be achieved in the synthetic spectrum by
including undetached Na~I~D lines.  To illustrate the necessity of
detaching the He~I lines, the bottom panel of Figure~6 shows how they
appear when they are undetached, i.e., when they are allowed to form
down to the photospheric velocity of 6000~\kms.  These synthetic
absorptions obviously are insufficently blueshifted.

The top panel of Figure~7 is like Figure~3 but with only the O~I
lines.  The optical depth of the reference line, \lam7773, is 1.  The
\lam7773 line accounts for at least most of an observed feature; in
some of the other observed spectra this feature may be partly produced
by Mg~II
\lam7890.  The O~I
\lam9264 line may be responsible for a weak observed feature, while
the \lam8446 feature usually is overwhelmed by the Ca~II infrared
triplet in SNe~Ib.  Whenever we use O~I lines in the synthetic spectra
of this paper, the optical depth of the reference line is near 1.

The bottom panel of Figure~7 is like Figure~3 but with only the Ca~II
lines.  The optical depth of the reference line, \lam3933, is 300.
Only the H\&K lines (\lam3933, 3968) and the infrared triplet
(\lam\lam8542,~8662,~8498) produce observable features, both of which
are very strong.  The notch in the synthetic spectrum near 8400~\AA\
nicely matches an observed feature.  In this synthetic spectrum the
Ca~II lines are detached to 7000~\kms\ to match the infrared triplet,
but in most of our synthetic spectra the Ca~II lines are undetached.

The top panel of Figure~8 is like Figure~3 but with only the Ti~II
lines.  The optical depth of the reference line, \lam4550, is 1.  The
Ti~II lines are used to help match the broad observed absorption
trough between 4100 and 4500~\AA.  We consider the presence of Ti~II
lines in the observed spectrum to be probable but not definite. [We do
consider them to be definite in peculiar subluminous SNe~Ia such as
SN~1991bg (Filippenko et~al. 1992) and SN~1999by (Garnavich
et~al. 2001), and in the Type~Ic SN~1994I (Millard et~al. 1999)].
Whenever we use Ti~II lines in this paper, the optical depth is near
1.  None of our other conclusions would be affected by omitting the
Ti~II lines.

The bottom panel of Figure~8 is like Figure~3 but with only the Sc~II
lines.  The optical depth of the reference line, \lam4247, is 0.5.
The Sc~II lines are used mainly to get a peak in the synthetic
spectrum near 5500~\AA.  The price to be paid is an overly strong
synthetic absorption produced by
\lam4247.  Sc~II lines are plausibly present in SNe~Ib 
because they are expected (in LTE) to appear at low temperatures
(Hatano et~al. 1999) and they appear in SNe~II.  Harkness
et~al. (1987) suggested that in SN~1984L the observed emission peak
near 5470~\AA\ was caused by the early emergence of blueshifted [O~I]
\lam5577 nebular--phase emission, but Swartz et~al. (1993) found
this to be unlikely for the Type~Ic SN~1987M.  For a discussion of the
possibility of an early emergence of blueshifted [O~I]
\lam5577 emission in the Type~IIb SN~1996cb, see Qiu et~al. (1999) and
Deng, Qiu, \& Hu (2001).  In our view the 5500~\AA\ emission in the
fiducial spectrum of SN~1999dn probably, but not definitely, is
produced by Sc~II lines.  Whenever we use them, the optical depth of
the reference line is near 1.  None of our other conclusions would be
affected by omitting the Sc~II lines.

Figure~9 is like Figure~3 except that the power--law index $n$ has
been reduced from 8 to 5 (and the optical depths of the reference
lines have been correspondingly reduced to keep the synthetic features
from becoming too strong).  Now the blue wings of the synthetic
absorptions produced by the Ca~II infrared triplet, Ca~II H\&K, and
He~I
\lam10830 fit better than in Figure~3, but the synthetic
absorptions produced by
\lam5876 and \lam6678 extend too far to the blue.  This
reflects the fact that in real supernovae, contrary to our assumption,
the line optical depths do not all follow the same power law
(or any power law).  A slower radial decline of the optical depth of
\lam10830 line, compared to the optical He~I lines, is expected on the
basis of the nonthermal--excitation calculations of Lucy (1991; his
Figure~3) and Swartz et~al. (1993; their Figure~11).  Note that if the
blue edge of the \lam10830 absorption really extends to 9000~\AA, as
it does in the synthetic spectrum of Figure~9, then the line is
forming all the way out to 50,000~\kms.  

Matheson et~al. (2001) demonstrated that the absorption produced by
\lam6678 (the lower level of which is 1s2p~$^1$P$^0$) becomes weaker
with time relative to the absorptions produced by \lam5876 and
\lam7065 (both 1s2p~$^3$P$^0$); this also is to be expected on the
basis of the results of Lucy and of Swartz et~al., because the singlet
resonance transitions to the ground state become less opaque as the
ejecta density decreases through expansion.

As mentioned above, Deng et~al. (2000) identified C~II lines in their
September~14 spectrum of SN~1999dn, obtained only three days before
our fiducial spectrum of September~17.  The identification of C~II
\lam6580 for the 6300~\AA\ absorption was supported by attributing a
weak absorption near 4580~\AA\ to C~II \lam\lam4738,~4745.  The
reasons that we don't introduce C~II lines for the fiducial spectrum
(apart from the fact that the 4580~\AA\ absorption doesn't appear
distinctly in the fiducial spectrum) are that (1) as mentioned above,
Fe~II lines could be responsible for the 6300~\AA\ absorption, and (2)
the absorption near 4580~\AA\ in the Beijing spectrum of September~14
might be produced by He~I \lam4731 (see the top panel of Figure~6)
and/or lines of Sc~II (see the bottom panel of Figure~8).

\subsection {SNe~2000H, 1999di, and 1954A: Hydrogen in SNe~Ib}

Now we turn to SN~2000H, an event that has conspicuous He~I lines but
that according to Benetti et~al. (2000) also has hydrogen lines.
Figure~10 compares the $+5$ day spectrum of SN~2000H with a synthetic
spectrum that has \vp=8000~\kms\ and
\tb=6500~K, and contains hydrogen lines in addition to the ions
used above for the fiducial spectrum of SN~1999dn.  The He~I lines are
detached at 9000~\kms, where the optical depth of the reference line
is 2.  The hydrogen lines are detached at 13,000~\kms\, where the
optical depth of the reference line, H$\alpha$, is 2.5.  With this
detachment velocity, H$\alpha$ accounts for at least most of the
6300~\AA\ absorption and H$\beta$ accounts for the unusual notch in
the emission peak near 4650~\AA.  A closer view of the H$\beta$ region
is provided in Figure~11.  The presence of an absorption that is
attributable to H$\beta$ provides strong support for the presence of
hydrogen in SN~2000H.

Matheson et~al. (2001) noted the presence of an unusually deep
6300~\AA\ absorption in SN~1999di, and mentioned that it could be
Si~II \lam6355, C~II
\lam6580, or H$\alpha$.  Si~II can now be rejected 
because its absorption would be blueshifted by only 2600~\kms, which
is too much lower than than our value of \vp=6000~\kms.  Figure~12
compares our earliest spectrum of SN~1999di, obtained 21 days after
maximum, with the $+19$ day spectrum of SN~2000H.  The similarity of
these two spectra is remarkable (apart from differences at wavelengths
longer than 9000~\AA\ where both spectra are noisy).  The narrow
H$\beta$ absorption of SN~2000H also can be seen in SN~1999di.
Figure~13 compares the $+21$ day spectrum of SN~1999di with a
synthetic spectrum that has \vp=7000~\kms\ and
\tb=4500~K and contains the same ions as Figure~10 for SN~2000H.  
Hydrogen is detached at 12,000~\kms.  Figure~14 shows a closer view of
the H$\beta$ region.

Could the 6300~\AA\ absorption in SNe~2000H and 1999di be produced by
C~II \lam6580\ rather than H$\alpha$?  We think not.  It would be
surprising to see a deep C~II \lam6580 absorption in SNe~Ib, when even
in SNe~Ic this line never forms a deep absorption and, if present at
all, is hard to identify.  It also would be surprising that the C~II
feature would be so detached in SNe~Ib. The top panel of Figure~15 is
like Figure~13 except that only C~II lines, detached at 13,000~\kms,
are used.  Although \lam6580 can account for the 6300~\AA\ absorption as
well as H$\alpha$ does, and
\lam\lam7236,~7231 are not a problem because, being more detached than
He~I, they fall near or within the strong feature produced by He~I
\lam7065, the absorption produced by C~II
\lam\lam4738,~4745 is much too strong (at least in LTE at 7000~K).

Could the 6300~\AA\ absorption in SNe~2000H and 1999di be produced by
Ne~I $\lambda$6402 rather than H$\alpha$?  No, it cannot.  The bottom
panel of Figure~15 shows that undetached Ne~I fails in two ways: (1)
the absorption produced by $\lambda$6402 is too far to the blue, and
(2) even though the synthetic absorption produced by $\lambda$6402 has
not been made strong enough, other unwanted features already are
present.

In view of these difficulties with C~II and Ne~I, and the apparent
presence of H$\beta$ in the observed spectra, we consider the
identification of hydrogen lines in SNe~2000H and 1999di to be
definite.

SN~1954A appears to have been spectroscopically akin to SNe~2000H and
1999di.  McLaughlin (1963) and Branch (1972) identified He~I lines in
photographic spectra of SN~1954A obtained at the Lick Observatory and
the Mount Wilson and Palomar Observatories, respectively. Consequently
SN~1954A usually has been regarded to have been a Type~Ib supernova,
but on occasion doubts has been expressed about the classification
because of the low quality of the photographic spectra compared to
modern observations.  Figure~16 compares microphotometer tracings of
the two earliest spectra of SN~1954A (Blaylock et~al. 2000), obtained
on blue-- and red--sensitive emulsions 46 days after maximum light,
with the $+19$ day spectrum of SN~2000H.  Only this one spectrum of
SN~1954 was obtained on a red--sensitive emulsion.  The SN~1954A
spectra are not actually relative flux but merely a measure of the
transmission through the photographic plate. Parts of the spectra were
overexposed so the features are distorted; in particular, emission
peaks tend to be suppressed.  The shapes of these spectra also are
strongly influenced by the wavelength dependences of the emulsion
sensitivities; e.g., neither emulsion is sensitive around 5200~\AA\
and the sensitivity falls off steeply between 6500 and 7000~\AA.
Nevertheless, some of the absorptions in SN~1954A can be located and
they correspond well with the absorptions in SN~2000H, including the
two attributed to H$\alpha$ and H$\beta$.  Branch (1972) considered
the possibility of H$\alpha$ and H$\beta$ absorptions in SN~1954A,
blueshifted by 10,800~\kms\ in the observer's frame; the value of $cz$
for the parent galaxy is now known to be 291~\kms, so the hydrogen
lines are blueshifted by about 11,000
\kms\ in the supernova frame. Branch also considered the possibility
of Ne~I lines in SN~1954A. This is perhaps still not ruled out, but
given the apparent resemblance of SN~1954A to SNe~2000H, we prefer the
hydrogen identification.

In terms of apparent magnitude, SN~1954A was the fourth brightest
supernova of the twentieth century, surpassed only by the Type~II
SN~1987A and the Type~Ia SNe~1972E and 1937C (Barbon et~al. 1999).
SN~1954A occurred in the star--bursting dwarf galaxy NGC~4214, at a
distance of only about 4~Mpc (Leitherer et~al. 1996), more than 10
times closer than SNe~2000H and 1999di, so it is more amenable to
studies of the environment in which it exploded.  We note that Van
Dyk, Hamuy, \& Filippenko (1996) found SN~1954A to be unusual among
SNe~Ib in that it was not near any visible H~II region, and that a
deep VLA search for radio emission at the site of SN~1954A carried out
by J.~Cowan and D.~Branch in May, 1986, resulted in a three--sigma
upper limit to the flux density at 20~cm of 0.068~mJy (Eck 1998),
which corresponds to a monochromatic luminosity of less than one
twentieth of Cas~A.

\subsection{Other Selected Comparisons}

Figure~17 compares a spectrum of SN~1984L obtained 9 days after
maximum with a synthetic spectrum that has
\vp=8000~\kms\ and \tb= 6500~K, and includes hydrogen lines 
detached at 15,000~\kms.  The Fe~II lines fit very well, and the fit
to the other features is satisfactory.  Given the presence of
H$\alpha$ in SNe~2000H, 1999di, and 1954A, we assume that the
6300~\AA\ absorption is produced by H$\alpha$.  However, because the
H$\alpha$ optical depth at the detachment velocity is only 0.6, there
is no support for this identification from H$\beta$ because it is too
weak to see. (The oscillator strength of H$\beta$ is about one fifth
that of H$\alpha$.)  Figure~18 compares the $+9$ day spectrum of
SN~1984L with the $+5$ day spectrum of SN~2000H.  The spectra are
similar except that in SN~1984L the 6300~\AA\ absorption is weaker and
the 4560~\AA\ absorption is not visible.  This shows that while our
assumption that the 6300~\AA\ absorption in SN~1984L is produced by
H$\alpha$ is reasonable, it is not proven.  It is conceivable that
C~II \lam6580 or (more plausibly because it wouldn't need to be
detached) Ne I $\lambda$6402 could be responsible for the 6300~\AA\
absorption in SN~1984L and other typical SNe~Ib, and be overwhelmed by
H$\alpha$ only in events such as SNe~2000H, 1999di, and 1954A.  We
proceed on the assumption that at early times the 6300~\AA\ absorption
is produced by H$\alpha$ in all the SNe~Ib of our sample.

As an example of a comparison at a earlier time when \vp\ is higher,
Figure~19 compares the earliest spectrum for which good wavelength
coverage is available, the Beijing spectrum of SN~1999dn 10 days
before maximum, with a synthetic spectrum that has
\vp=14,000~\kms\ and \tb= 6500~K.  Now hydrogen lines are detached at
18,000~\kms, where the H$\alpha$ optical depth is 1.3.  The fit is
good, except near 6600~\AA.

Figure~20 compares a spectrum of SN~1998dt obtained 32 days after
maximum with a synthetic spectrum that has \vp = 9000~\kms\ and \tb=
5000~K.  This comparison is shown because, as will be seen below,
although the spectrum has a typical SN~Ib appearance the inferred
value of
\vp\ is unusually high for a SN~Ib this long after maximum.  The 
fit to the Fe~II features is unusually poor, but this value of \vp\
does give the best fit in the 4300 to 5300~\AA\ region, and a
significantly lower value would give a noticeably worse fit.  (The
narrow H$\alpha$ emission from an H~II region in the parent galaxy is
very close to 6563~\AA, which shows that the high required value of
\vp\ is not due to the observed spectrum having been inadvertently
overcorrected for parent galaxy redshift.)

Now we briefly consider the events of our sample for which the times
of maximum light are unknown.  The spectrum of SN~1996N indicates that
it was discovered not long after maximum.  Because we use hydrogen
lines in the synthetic spectrum for SN~1996N, the comparison with the
observed spectrum is shown in Figure~21.  Helium lines are undetached
and the optical depth of the reference line is 5; hydrogen lines are
detached at 17,000~\kms\ where the optical depth of H$\alpha$ is 0.5.
Overall, the fit is good.

SNe~1991ar and 1997dc do not appear to be unusual provided that they
were discovered well after maximum light.  As discussed by Matheson
et~al. (2001), the spectra of SN~1998T are seriously contaminated by
light from the parent galaxy; taking this into consideration, SN~1998T
does not appear to be unusual provided that it was discovered about a
week after maximum.  As far as we can tell, SNe~1991ar, 1997dc, and
1998T were typical SNe~Ib, but they were not observed early enough to
check on the presence of H$\alpha$.

\section{RESULTS AND DISCUSSION}

The most important of the fitting parameters that have been used for
the synthetic spectra are collected in Table~2.  The spectra are
listed in order of time with respect to maximum light so only the
supernovae for which we have an estimate of the time of maximum light
appear in the table.  (The date 0703, for example, refers to July~3.)

For the six SNe~Ib for which we have estimates of the time of maximum
light, \vp\ is plotted against time in Figure~22. The tightness of the
relationship is striking.  SN~1998dt at 32 days after maximum seems to
stand out; otherwise, the scatter about the mean curve is about what
should be expected from our nominal errors of
1000~\kms\ in \vp\ and a few days in the dates of maximum light.  For
the simple case of constant opacity and a $v^{-n}$ density
distribution, the velocity at the photosphere would decrease with time
as $v_{phot}
\propto t^{-2/(n-1)}$.  The line in Figure~22, the best power--law fit
to the data (excluding SN~1998dt at 32 days), corresponds to $n=3.6$.
Considering that the opacity is not really constant, that the actual
density distribution does not really follow a single power law over a
wide velocity range, and that the best power--law index for fitting
the spectra is not well constrained, not much significance should be
attched to the difference between $n=3.6$ and $n=8$.

Our adopted values of \vp\ can be used to make rough estimates of the
mass and kinetic energy above the photosphere.  For spherical symmetry
and a $v^{-n}$ density distribution, the mass (in $M_\odot$) and the
kinetic energy (in 10$^{51}$ ergs) above the electron--scattering
optical depth $\tau_{es}$ are (Millard et~al. 1999)

$$ M=1.2 \times 10^{-4}\ v_4^2\ t_d^2\ \mu_e\ {{n-1}\over{n-3}}\ \tau_{es},
\eqno (1) $$

$$ E=1.2 \times 10^{-4}\ v_4^4\ t_d^2\ \mu_e\ {{n-1}\over{n-5}}\ \tau_{es},
\eqno (2) $$

\noindent where $v_4$ is \vp\ in units of 10,000~\kms, $t_d$ is the time since
explosion in days, $\mu_e$ is the mean molecular weight per free
electron, and the integration is carried out to arbitrarily high
velocity.  If we assume that maximum light occurs 20 days after
explosion, that $n=8$, that $\mu_e = 8$ (e.g., half--ionized helium or
singly ionized oxygen), and that \vp\ is at $\tau_{es}=1$, then at
maximum light
\vp=10,000~\kms\ (Figure~22) gives $M=0.5~M_\odot$ and $E=0.9 \times
10^{51}$ ergs.  At 20 days after maximum, using \vp=7000~\kms\ and
keeping the other parameters the same gives $M=1.1~M_\odot$ and $E=0.9
\times 10^{51}$ ergs.  In reality, of course, the kinetic energy above
the 7000~\kms\ photosphere must be greater than that above the
10,000~\kms\ photosphere. If we use $n=4$ instead of $n=8$ between
7000 and 10,000~\kms\, then we obtain a total of $M=1.5~M_\odot$ and
$E=1.4 \times 10^{51}$ ergs above the 7000~\kms\ photosphere.  

Figure~22 provides some constraints on models of SNe~Ib.  The
hydrodynamics must account for the velocity at the photosphere as a
function of time, and the ensemble of SN~Ib progenitors must be
consistent with the tightness of the relationship.  The small scatter
suggests that the masses and the kinetic energies of the SNe~Ib of our
sample are similar, and it does not leave much room for the influence
on \vp\ of departures from spherical symmetry.

Figure~23 shows the minimum velocities of the He~I lines (squares when
undetached and diamonds when detached).  There appears to be a
standard pattern (again with the possible exception of SN~1998dt at 32
days).  Before and near the time of maximum the He~I lines tend to be
undetached.  After maximum the lines tend to be detached, but the
detachment velocities tend to decrease with time, from about 10,000 to
7000~\kms. This means that the fraction of helium in this velocity
range that is in the lower levels of the optical He~I lines is
increasing with time faster than $t^2$.  (The matter density is
decreasing as $t^{-3}$ by expansion but the Sobolev optical depth also
is proportional to $t$ because it is inversely proportional to the
velocity gradient.)  The increasing fraction of helium in the excited
levels may be understandable in terms of the decreasing column depth
between the nickel core and the helium layers, and the decreasing
detachment velocity may mean that the fractional helium abundance is
lower at lower velocities.  In any case, some helium is present at
least down to 7000~\kms.  Our estimate above for the total mass above
the 7000~\kms\ photosphere was 1.1 to 1.5~M$_\odot$, which is a rough
upper limit on the mass of helium above 7000~\kms.  There could be
more helium below 7000~\kms.

Figure~23 provides more constraints on models of SNe~Ib.  The radial
profile of the helium abundance, together with that of the $^{56}$Ni
that is responsible for exciting it, should account for the He~I
velocities and optical depths (Table~2).

Figure~23 also shows the minimum velocities of the hydrogen lines
(circles), which always are detached.  In the three events for which
we are convinced of the hydrogen identifications --- SNe~2000H,
1999dn, and 1954A --- the minimum hydrogen velocity is between 11,000
and 13,000~\kms.  In the events in which we assume H$\alpha$ to be
present at early times --- SNe~1983N, 1984L, 1999dn, and 1996N (the
latter is not shown in Figure~23 because the date of maximum light is
unknown) --- the detachment velocities tend to decrease with time but
they are consistent with similar minimum velocities of the hydrogen.
Thus the available evidence is consistent with the proposition that
SNe~Ib in general have hydrogen down to $11,000 - 13,000$~\kms.  [For
comparison, in the Type~IIb SN~1993J the characteristic velocity of
the ejected hydrogen was about 9000~\kms\ (Patat et~al. 1995; Utrobin
1996; Houck \& Fransson 1996), and in the Type~IIb SN~1996cb it was
about 10,000~\kms\ (Deng et~al. 2001).]  A challenge for those who
study the complicated evolution of massive stars in binary systems
(e.g., Podsiadlowski, Joss, \& Hsu 1992; Nomoto, Iwamoto, \& Suzuki
1995; Wellstein, Langer,
\& Braun 2001) is to understand why many or perhaps even all stellar
explosions that develop strong He~I lines should eject at least a
small amount of hydrogen.

The hydrogen mass that is required to give an H$\alpha$ optical depth
of unity depends on the fraction of hydrogen that is in the Balmer
level.  In LTE, with an electron density of $10^9$~cm$^{-3}$, the
Balmer fraction peaks around $10^{-9}$ near 6000~K.  In this case we
estimate that a hydrogen mass on the order of $10^{-2}~M_\odot$ would
be required.  Non--LTE calculations that take nonthermal excitation
into account are needed for a more reliable estimate of the hydrogen
mass.

The optical depths of the helium lines, and especially the hydrogen
lines, are not very high, even though the corresponding absorption
features are distinct and fairly deep.  This reflects a simple
geometric aspect of supernova line formation: as explained in Jeffery
\& Branch (1990), absorption features formed by detached lines are
deeper than those formed by undetached lines.  This point is
illustrated in Figure~24, which shows that when hydrogen is detached
from the photosphere by a factor of two and H$\alpha$ has $\tau=2$,
its absorption feature is deeper than that of an undetached H$\alpha$
that has $\tau=10$.  The H$\alpha$ optical depths in our synthetic
spectra for SNe~2000H, 1999di, and 1954A are not high, so if they were
only mildly lower these events would look like typical
SNe~Ib. Similarly, moderately lower He~I line optical depths would
transform a SN~Ib into a SN~Ic.

Figure~24 also illustrates how undetached hydrogen lines are more
``obvious'' than detached lines.  First, undetached lines have
conspicuous narrow, rounded emission peaks while detached lines have
inconspicuous broad, flat peaks.  Note also that although the
H$\alpha$ absorption is deeper in the detached spectrum, the H$\beta$
absorption is deeper in the undetached spectrum.  This is because in
the undetached spectrum the optical depth at the photosphere of
H$\beta$ is about 2 while in the detached spectrum the optical depth
at the detachment velocity is only about 0.4.  An optical depth as low
as 0.4 can produce only a shallow absorption, even when the line is
detached.  For these reasons, supernovae that have undetached hydrogen
lines have obvious hydrogen lines and are classified as Type~II.
Supernovae that have detached hydrogen lines are classified as Type~Ib
because the presence of hydrogen is not immediately obvious, even when
the H$\alpha$ absorption is as deep as it is in SNe~2000H, 1999di, and
1954A.  SNe~IIb are those that have undetached hydrogen lines when
they are first observed.  In some cases, whether an event is
classified as Ib or IIb may depend on how early the first spectrum is
obtained.

The implication of the previous paragraphs is that the spectroscopic
differences between SNe~IIb, the SNe~Ib that have deep H$\alpha$
absorptions, and typical SNe~Ib may be caused mainly by mild
differences in the hydrogen mass.  For a given kinetic energy, the
lower the hydrogen mass the higher the minimum velocity of the ejected
hydrogen.  Similarly, the spectroscopic differences between typical
SNe~Ic and SNe~Ib could be caused mainly by moderate differences in
the helium mass.  For example, Matheson et~al. (2001) found higher
blueshifts of the O~I $\lambda$7773 line in SNe~Ic than in SNe~Ib.
For a given kinetic energy, the lower the helium mass the higher the
minimum velocity of the ejected helium, and therefore the higher the
velocity of the ejected oxygen.  These suggestions are not original to
this paper, but they are strengthened by our finding that the H~I and
He~I optical depths in SNe~Ib are not very high.  These suggestions
also are not inconsistent with arguments, based on light curves, for
the existence of different physical classes of hydrogen--poor events
that cut across the conventional spectroscopic types (e.g.,
Clocchiatti \& Wheeler 1997).

The number of SNe~Ib for which good spectral coverage is available is
still relatively small.  More events should be observed to explore the
degree of the spectral homogeneity and to find out whether there is a
continuum of hydrogen line strengths.  Also needed are detailed
non--LTE spectrum calculations for supernova models having radially
stratified compositions, --- to determine the the hydrogen and helium
masses and the distribution of the $^{56}$Ni that is required to
excite the helium.  The possibility that nonthermally excited Ne~I can
produce spectral features strong enough to be seen needs to be
investigated.  Detailed non--LTE calculations for parameterized SN~Ib
models, using the PHOENIX code (e.g., Baron et~al. 1999), are
underway.

\bigskip

This material is based upon work supported by the National Science
Foundation under Grants No. AST--9986965 and AST--9731450 at Oklahoma
and AST--9987438 at Berkeley.  A.V.F. is grateful to the Guggenheim
Foundation for a Fellowship.

\clearpage

\begin {references}

\reference{} Barbon, R., Buondi, V., Cappellaro, E., \&
Turatto,~M. 1999, A\&AS, 139, 531

\reference{} Baron, E., Branch, D., Hauschildt, P. H.,
Filippenko,~A.~V., \& Kirshner,~R.~P. 1999, ApJ, 527, 739

\reference{} Benetti, S., Cappellaro, E., Turatto, M., \&
Pastorello,~A. 2000, IAU Circ. 7375

\reference{} Blaylock, M., Branch, D., Casebeer, D., Millard, J.,
Baron,~E., Richardson,~D., \& Ancheta,~C. 2000, PASP, 112, 1439

\reference{} Branch, D. 1972, A\&A, 16, 247

\reference{} Branch, D. 2001, in Supernovae and Gamma--Ray Bursts: The
Largest Explosions in the Universe, ed. M.~Livio (Cambridge: Cambridge
University Press), in press

\reference{} Casebeer, D., Branch, D., Blaylock, M., Millard, J.,
Baron,~E., Richardson,~D., \& Ancheta,~C. 2000, PASP, 112, 1433

\reference{} Clocchiatti, A. \& Wheeler, J. C. 1997, in Thermonuclear
Supernovae, ed. P.~Ruiz--Lapuente, R.~Canal, \& J.~Isern (Dordrecht:
Kluwer), p. 863

\reference{} Deng, J. S., Qiu, Y. L., Hu, J. Y., Hatano~K., \&
Branch~D. 2000, ApJ, 540, 452

\reference{} Deng, J., Qiu, Y., \& Hu, J. 2001, preprint

\reference{} Eck, C. R. 1998, PhD Thesis, University of Oklahoma

\reference{} Filippenko, A. V. 1997, ARAA, 35, 309

\reference{} Filippenko, A. V., et~al., 1992, AJ, 104, 1543

\reference{} Fisher, A. 2000, PhD Thesis, University of Oklahoma

\reference{} Garnavich, P. M., et~al., 2001, ApJ, submitted

\reference{} Garnavich, P., Challis, P., Jha, S., \& Kirshner,~R. 
2000, IAU Circ. 7366

\reference{} Germany, L., Schmidt,~B., Stathakis,~R., \& Johnston,~H.  
2000, IAU Circ. 6351

\reference{} Harkness, R. P., et al., 1987, ApJ, 317, 355

\reference{} Hatano, K., Branch, D., Fisher, A., Millard,~J., \&
Baron,~E. 1999, ApJS, 121, 233

\reference{} Houck, J. C., \& Fransson, C. 1996, ApJ, 456, 811

\reference{} Jeffery, D. J., \& Branch, D. 1990, in Supernovae,
ed. J.~C.~Wheeler, T.~Piran, \& S.~Weinberg (Singapore: World
Scientific), p. 149

\reference{} Kurucz, R. L. 1993, CDROM No. 1: Atomic Data for Opacity
Calculations, Cambridge, Smithsonian Astrophysical Observatory

\reference{} Leitherer, C., Vacca, W. D., Conti, P. S.,
Filippenko,~A.~V., Robert,~C., \& Sargent,~W.~L.~W. 1996, ApJ, 465,
717

\reference{} Lucy, L. B. 1991, ApJ, 383, 308

\reference{} Matheson, T., Filippenko, A. V., Li, W., Leonard, D. C., \&
Shields,~J.~C. 2001, AJ, 121, 1648

\reference{} McLaughlin, D. B. 1963, PASP, 75, 133

\reference{} Millard, J., et al., 1999, ApJ, 527, 746

\reference{} Nomoto, K., Iwamoto, K., \& Suzuki, T. 1995, Phys. Rep.,
256. 173

\reference{} Pastorello, A., Altavilla, G., Cappellaro, E., \&
Turatto,~M. 2000, IAU Circ. 7367

\reference{} Patat, F., Chugai, N., \& Mazzali, P. A. 1995, A\&A, 299, 715

\reference{} Podsiadlowski, Ph., Joss, P. C., \& Hsu, J. J. L. 1992,
ApJ, 391, 246

\reference{} Qiu, Y., Li, W., Qiao, Q., \& Hu,~J. 1999, AJ, 117, 736

\reference{} Richter, T., \& Sadler, E. M. 1983, A\&A, 128, L3

\reference{} Swartz, D. A., Filippenko, A. V., Nomoto, K., \&
Wheeler,~J.~C. 1993, ApJ, 411, 313

\reference{} Tsvetkov, D. Yu. 1987, Sov. Astron. Lett., 13, 376

\reference{} Utrobin, V. P. 1996, A\&A, 306, 219

\reference{} Van Dyk, S., Hamuy, M., \& Filippenko, A. V. 1996, AJ,
111, 2017

\reference{} Wellstein, S., Langer, N., \& Braun, H. 2001, A\&A, 369, 939

\reference{} Wheeler, J. C., Harkness, R. P., Clocchiati,~A.,
Benetti,~S., Brotherton,~M.~S., DePoy,~D.~L., \& Elias,~J. 1994, ApJ,
436, L135

\reference{} Woosley, S. E., \& Eastman, R. G. 1997, in Thermonuclear
Supernovae, ed. P.~Ruiz--Lapuente, R.~Canal, \& J.~Isern (Dordrecht:
Kluwer), p. 821

\end{references}

\clearpage

\begin{deluxetable}{lrrcrr} \tablenum{1} \tablewidth{0pc}
\tablecaption{Type Ib Supernovae} 

\tablehead{ \colhead{SN} & \colhead{Galaxy} & \colhead{$cz$ (km
s$^{-1}$)} & \colhead{t$_{max}$}}

\startdata

1954A & NGC 4214 & 291 & April 21 \\ 

1983N & NGC 5236 & 513 & July 17\\

1984L & NGC 991 & 1534 &  August 22 \\ 

1991ar & IC 49 &  *4520  &  *September 2 \\

1996N & NGC 1398 & 1491  &  *March 12 \\

1997dc & NGC 7678 & 3480 & *August 5 \\

1998T & NGC 3690 & *3080 & *March 3\\

1998dt & NGC 945 & *4580  &  September 12\\

1999di & NGC 776 & *4920  &  July 27 \\

1999dn & NGC 7714 & 2700   & August 31 \\ 
 
2000H & IC 454 & 3894  &  February 11  \\

\enddata
\end{deluxetable}

\clearpage

\begin{deluxetable}{lrrcrrrr} \tablenum{2} \tablewidth{0pc}
\tablecaption{Synthetic Spectrum Parameters} 

\tablehead{ \colhead{SN} & \colhead{date} & \colhead{epoch} &
\colhead{$v_{phot}$} & \colhead{$\tau$(He~I)} &
\colhead{$v_{det}$(He~I)} & \colhead{$\tau$(H)} &
\colhead{$v_{det}$(H)}}

\startdata

1983N & 0703 & -14 & 17000 &&&& \\

1983N &0706 &-11 & 13000 &&&&\\

1999dn& 0821& -10 & 14000   &   2.0 & 14000  & 1.5& 18000 \\

1983N & 0713 & -4 & 11000 &&&&\\

1983N & 0717 &  0 & 11000 &  2.5& 11000 & 0.8 & 15000 \\

2000H & 0211 &  0 & 11000 &  1 & 11000 & 6.0 & 13000 \\

1999dn &0831 &  0 & 10000    &  5.0 &11000 & 1.0 & 14000 \\

1983N & 0719 &  2 & 10000  & 4.0 &  10000&  0.7 & 14000 \\

2000H & 0216 &  5   &  8000    &     2.0 & 9000  &  2.5 & 13000 \\

1984L&  0830 &  8  & 9000    &  2.0 & 10000 &  0.6 & 15000 \\

1998dt& 0920 &  8 & 9000  &  4.0 &11000 &&\\

1984L & 0831&   9 &  8000  &   1.5 &  10000 &  0.6& 15000  \\

1983N & 0727 & 10 &  7000 & 7.0  & 8000 & 0.6 & 12000 \\

1999dn & 0910 & 10  &  7000  &  3.0 & 8000  &&\\

1984L & 0903 & 12   & 7000   &  1.0&  9000 & 0.5& 14000 \\

1999dn& 0914  &14  & 6000  & 10  &  7000       && \\

1999dn &0917  &17   &6000  &   10 &   8000 && \\

2000H &0302  &19    &6000   &   5.0&   8000&   2.0& 13000 \\

1999di &0817  &21  &7000  &  10 &8500  & 4.0 &  12000  \\

1984L &0919  &28    &5000 &&&& \\   

2000H &0313  &30   &5000   &   3.0&  7000& 1.5 &13000 \\

1984L &0923  &32    &5000   &  5.0&  7000  &&           \\         

1998dt &1015 & 33  &9000   & 10 &  9000 &&\\

1984L &0928  &37   &5000   & 10  & 6000 &&\\

1999dn &1008  &38 &   6000  &  10 &   7000&& \\ 

1999di &0910   & 45& 6000    & 10 & 7000 &   2.0 &  12000  \\

2000H &0330  &47   &5000      & 1.0&  7000 & 0.5& 12000 \\

1999di &0917  &52  &6000    & 10 & 7000& 1.0& 12000  \\

2000H &0408  &56   &4000                 &&&&   \\

1984L &1018  &57   &4000 &&&&\\

\enddata
\end{deluxetable}

\clearpage

\begin{figure}
\plotone{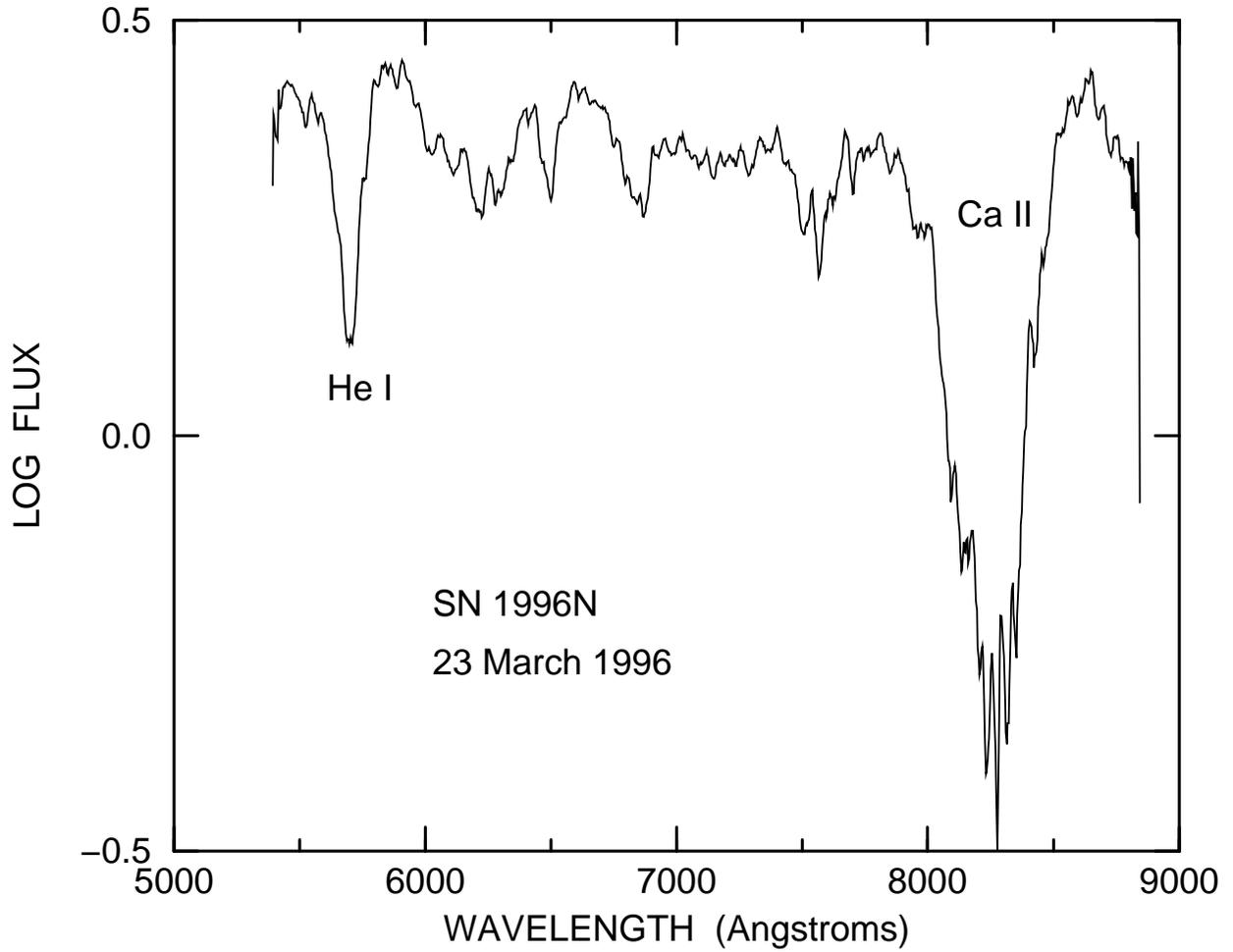}
\figcaption{A spectrum of SN~1996N obtained at the Anglo--Australian
Observatory by Germany et~al. (1999) on 1996 March~23, 11 days after
discovery.  The flux is in units of $10^{-17}$
erg~cm$^{-2}$~s$^{-1}$~\AA$^{-1}$.  All spectra displayed in this
paper have been deredshifted.}
\end{figure}

\begin{figure}
\plotone{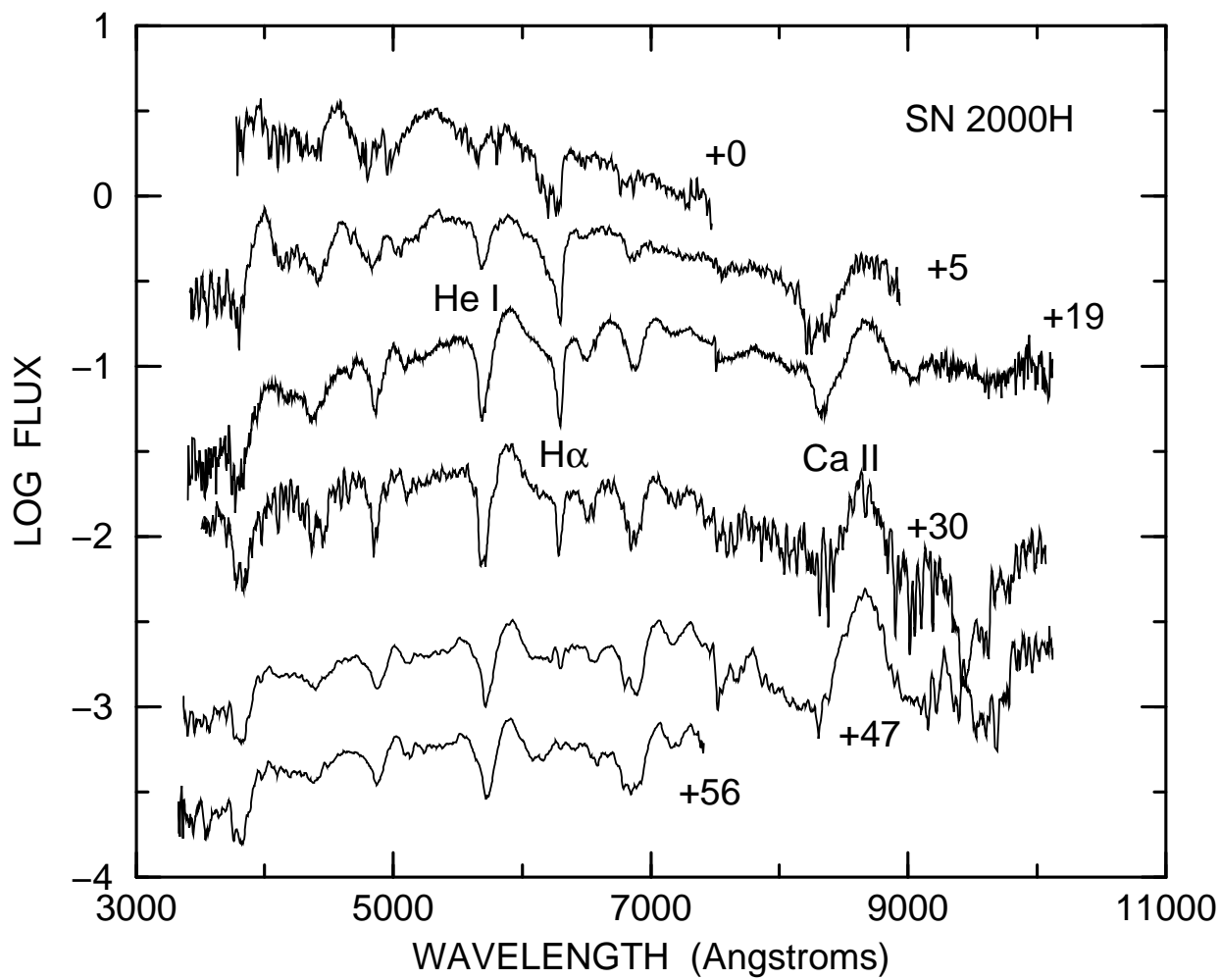}
\figcaption{Spectra evolution of SN~2000H. The times are in days after
2000 February~11. The flux is in units of $10^{-16}$
erg~cm$^{-2}$~s$^{-1}$~\AA$^{-1}$. The ordinate refers to the first
spectrum ($+0$ days) and the other spectra have been shifted downward
by 0.6, 1.1, 1.4, 2.6, and 2.9 dex.  The spectra were obtained as
follows: $0$ days, Asiago 1.82m+AFOSC, res. 20 \AA; $+5$ days,
Danish1.54m+DFOSC, res. 11 \AA; $+19$ days, ESO3.6m+EFOSC2, res. 15
\AA; $+30$ days (average of two spectra taken on two consecutive
nights), Danish1.54m+DFOSC, res. 11 \AA; $+47$ and $+56$ days,
ESO3.6m+EFOSC2, resol. 15 \AA. The telluric absorptions have been
removed.}
\end{figure}

\begin{figure}
\plotone{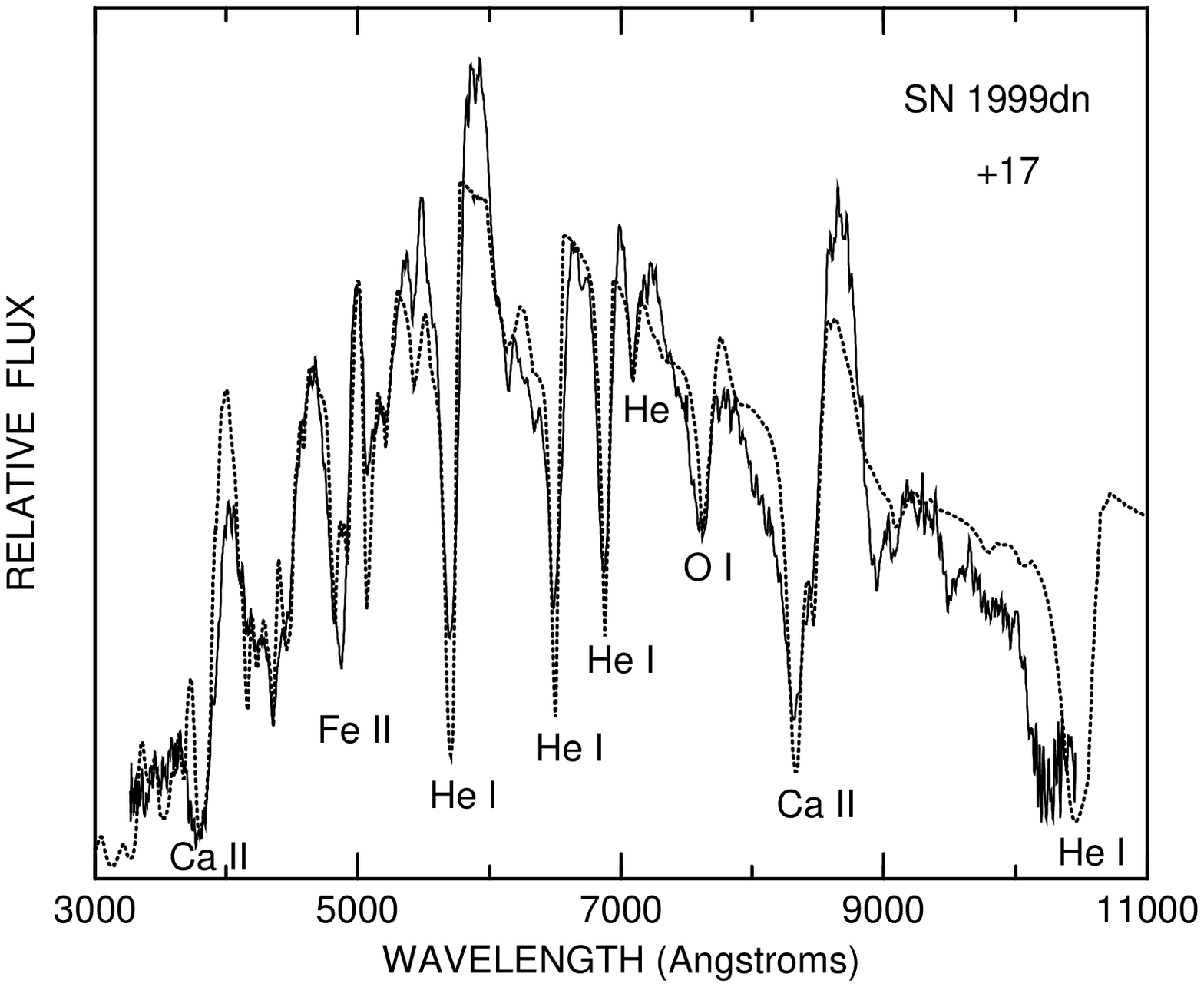}
\figcaption{The $+17$ day spectrum of SN~1999dn (solid line) is
compared with a synthetic spectrum (dotted line) that has $v_{phot}=
6000$~\kms\ and $T_{bb}=4800$~K, and contains lines of He~I, O~I,
Ca~II, Sc~II, Ti~II, and Fe~II.}
\end{figure}

\begin{figure}
\plotone{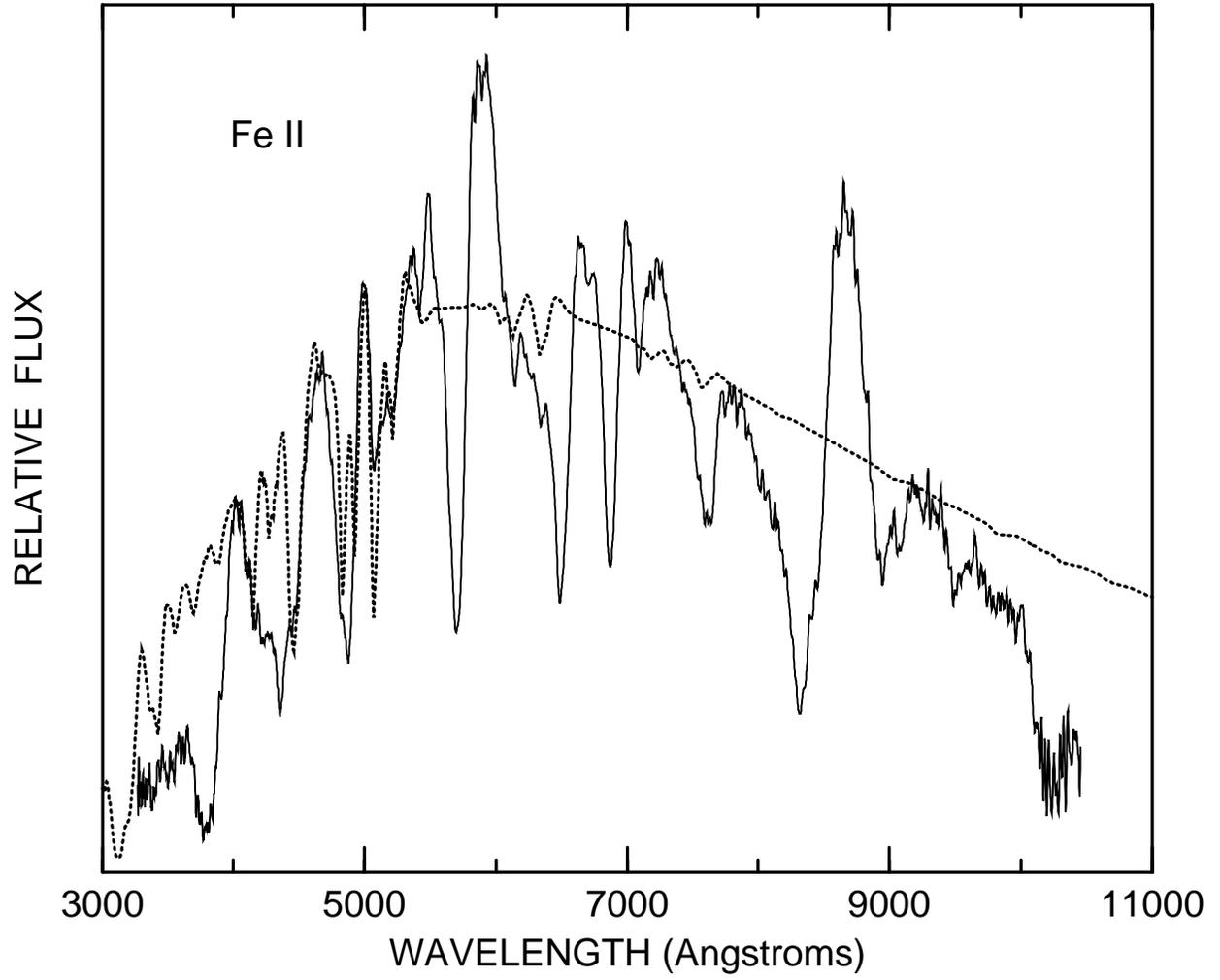}
\figcaption{Like Figure~3, but with only the Fe~II lines in the 
synthetic spectrum.}
\end{figure}

\begin{figure}
\plotone{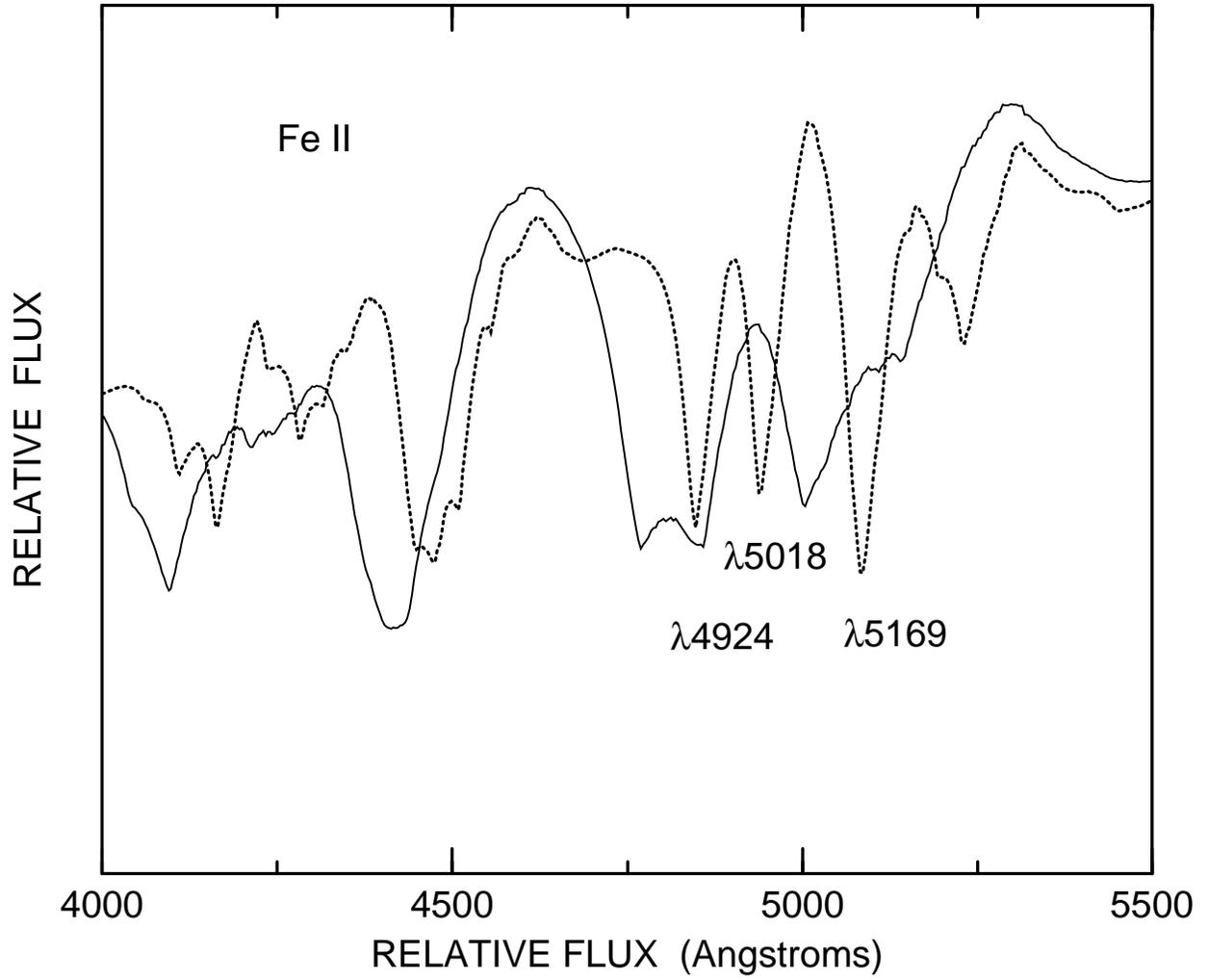}
\figcaption{A comparison of Fe~II synthetic
spectra having $v_{phot} = 5000$ (dotted line) and 10,000 \kms\ (solid
line).}\end{figure}

\begin{figure}
\plotone{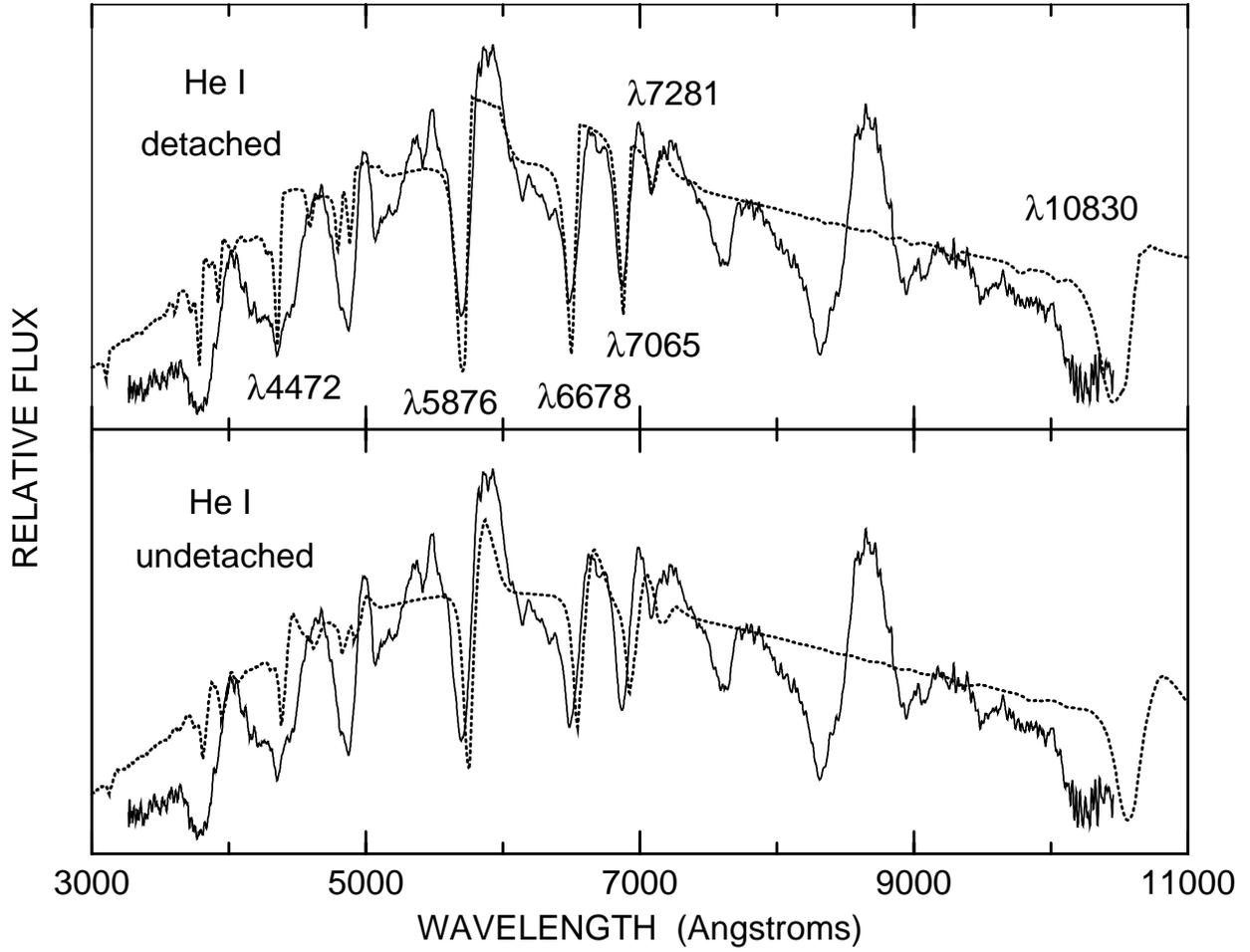}
\figcaption{Like Figure~3, but with only the He~I lines in the
synthetic spectra, detached at 8000
\kms\ ({\sl top panel}) and undetached ({\sl bottom panel}).}\end{figure}
\begin{figure}
\plotone{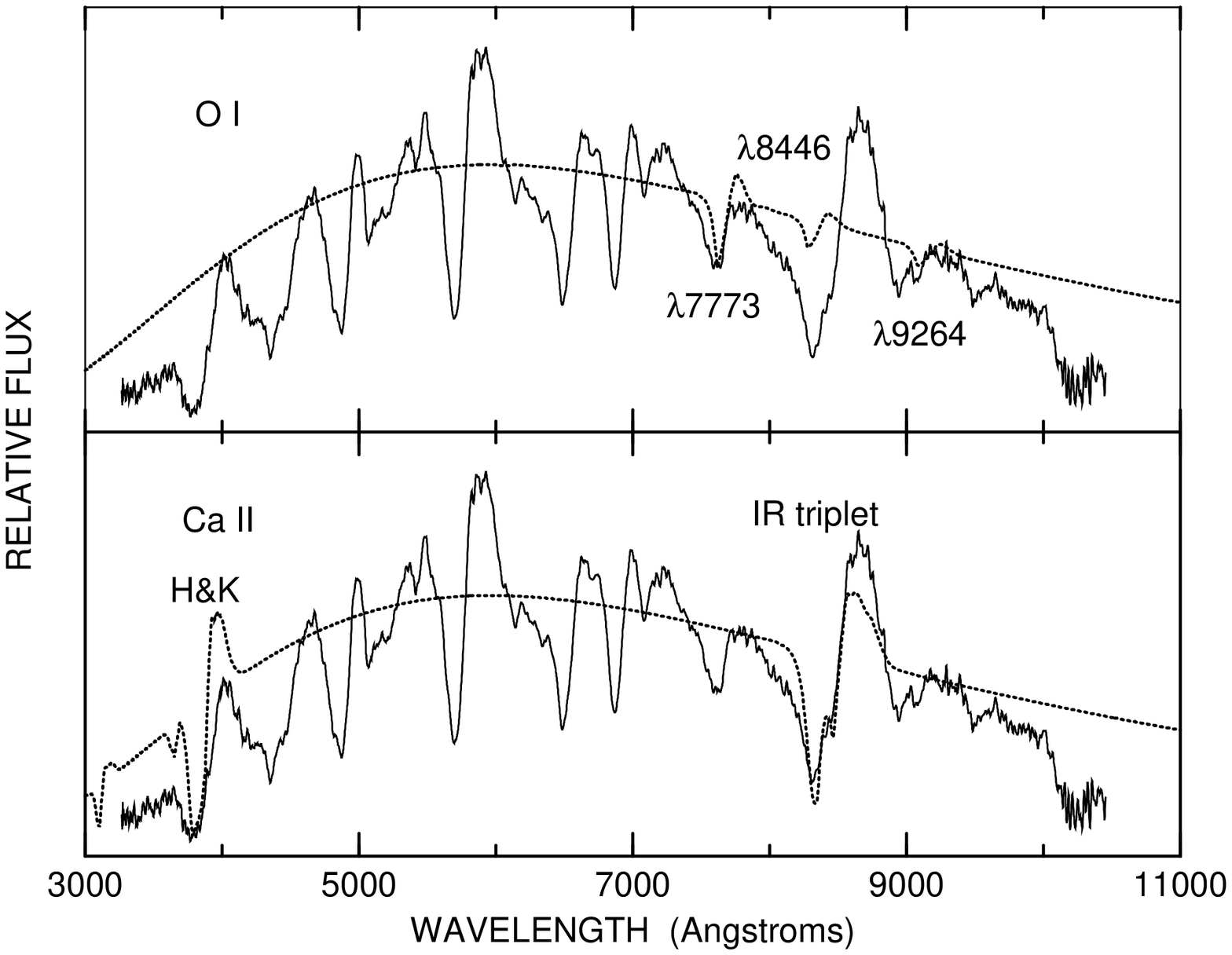}
\figcaption{Like Figure~3, but with only the O~I lines ({\sl top
panel}) and the Ca~II lines ({\sl bottom panel}) in the synthetic
spectra.}\end{figure}
\begin{figure}
\plotone{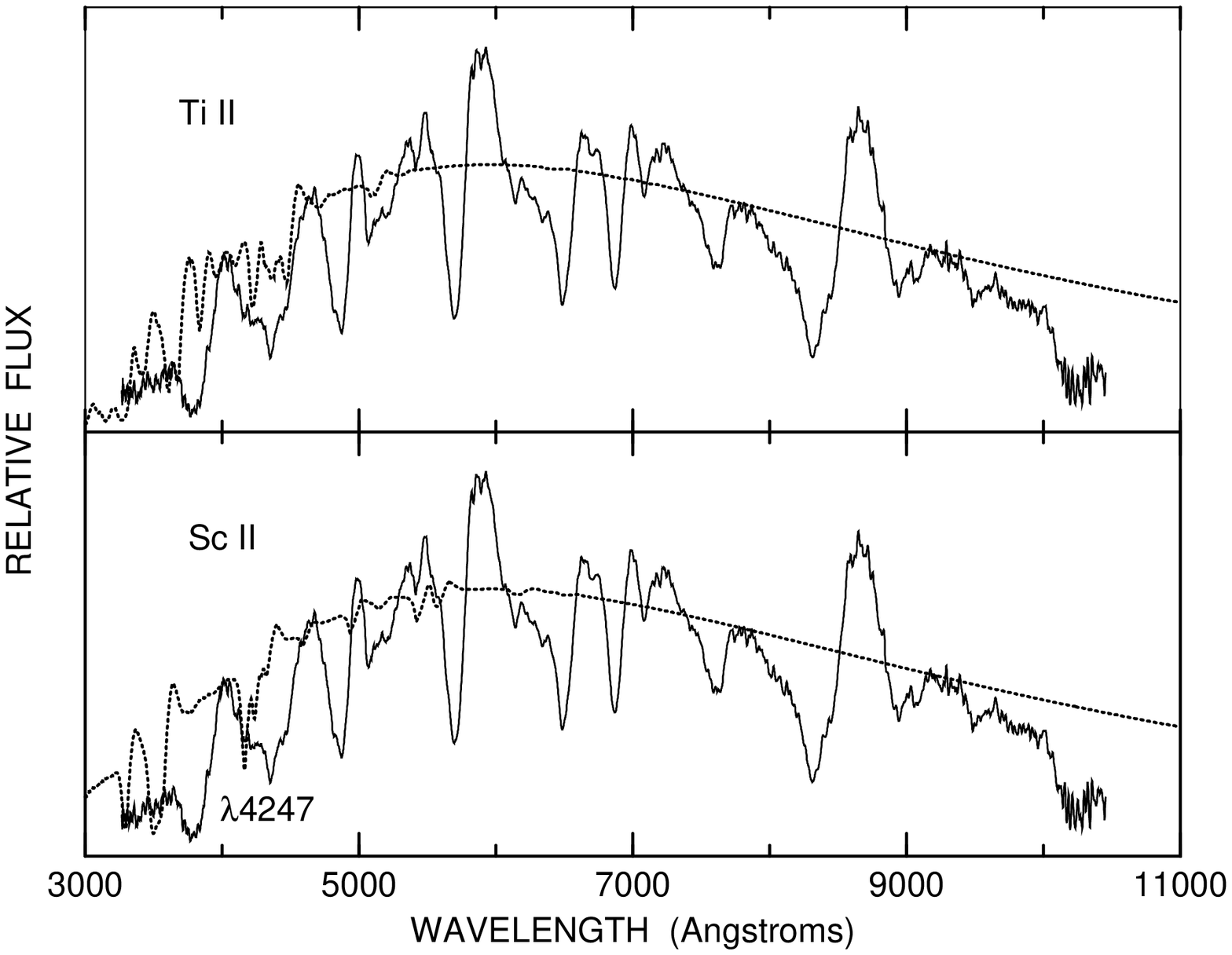}

\figcaption{Like Figure~3, but with only the Ti~II lines ({\sl top
panel}) and the Sc~II lines ({\sl bottom panel}) in the synthetic
spectra.}\end{figure}
\begin{figure}
\plotone{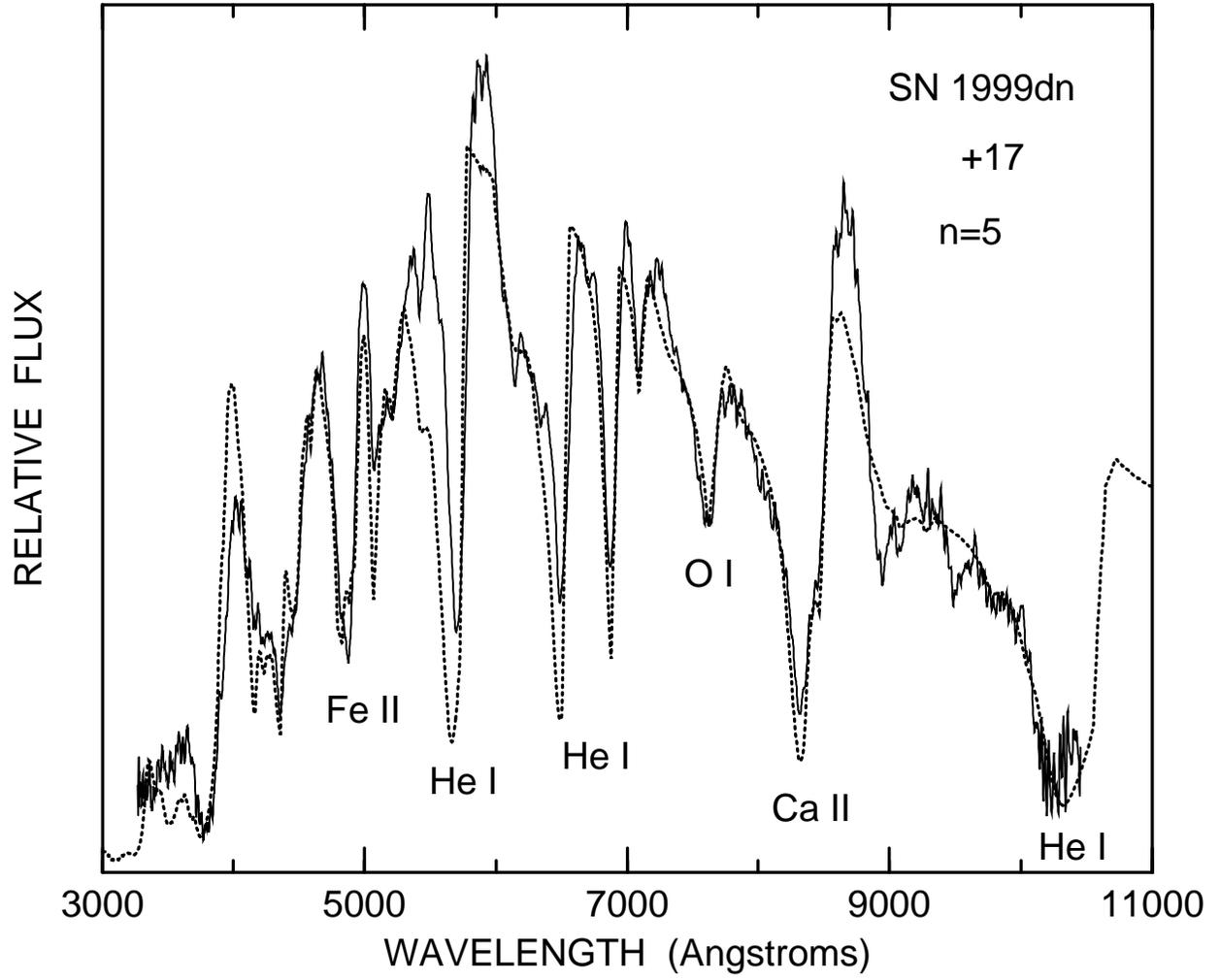}
\figcaption{Like Figure~3, but with $n=5$ instead of $n=8$.}\end{figure}
\begin{figure}
\plotone{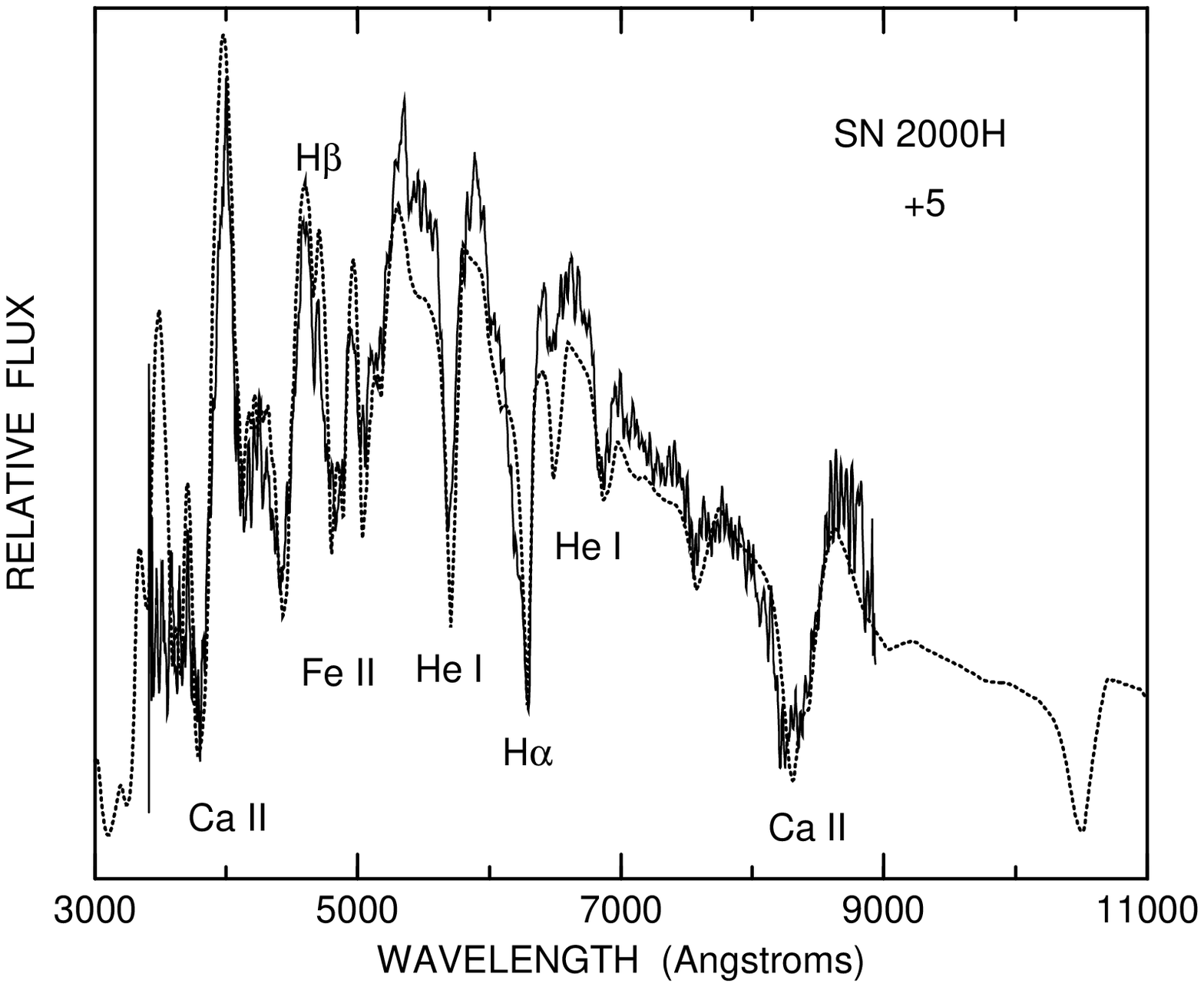}
\figcaption{The $+5$ day spectrum of SN~2000H (solid line) is
compared with a synthetic spectrum (dotted line) that has $v_{phot}=
8000$~\kms\ and $T_{bb}=6500$~K, and contains lines of H, He~I, O~I,
Ca~II, Sc~II, Ti~II, and Fe~II.}\end{figure}
\begin{figure}
\plotone{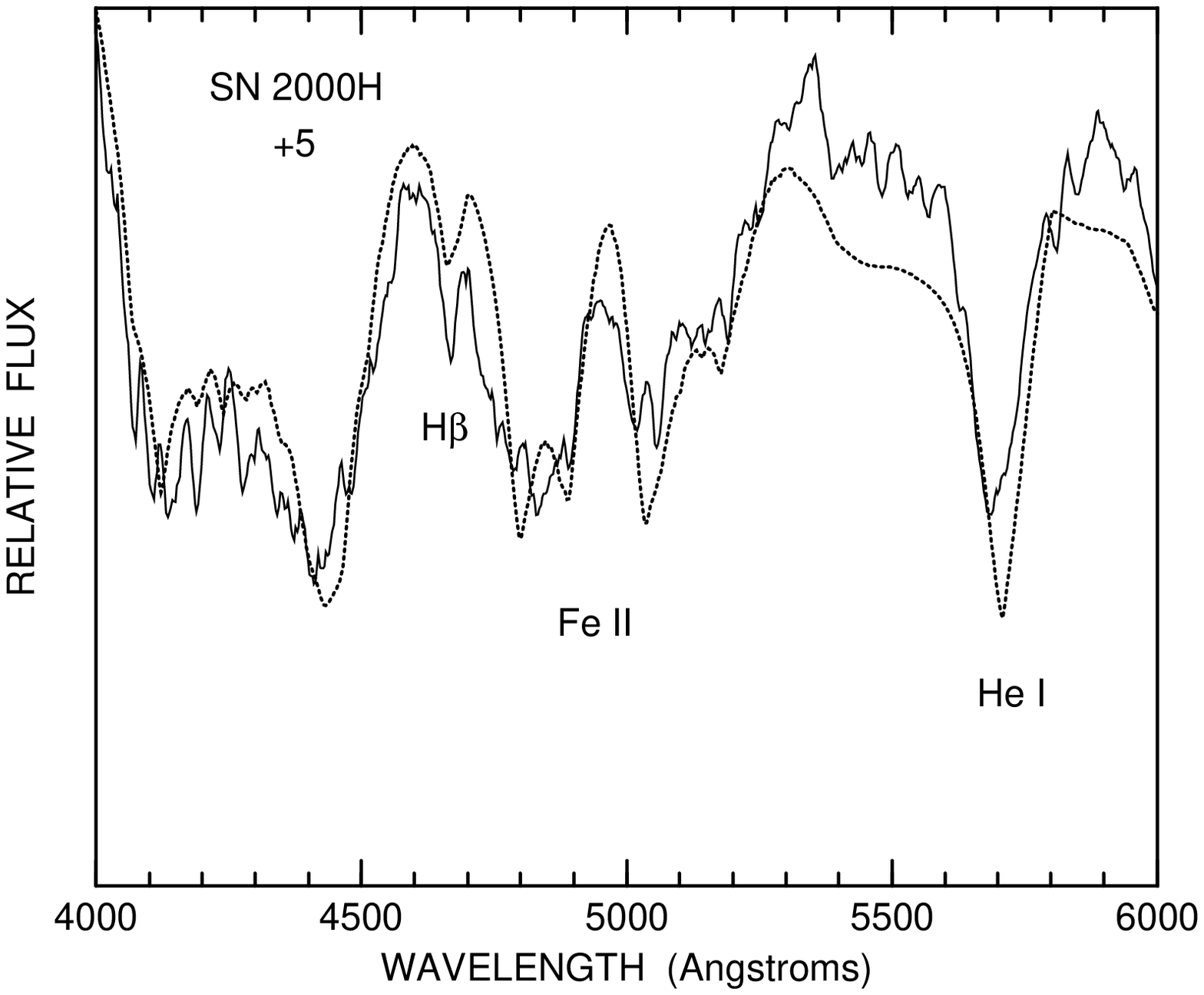}
\figcaption{Like Figure~10, but a better view of the H$\beta$ region.}\end{figure}

\begin{figure}
\plotone{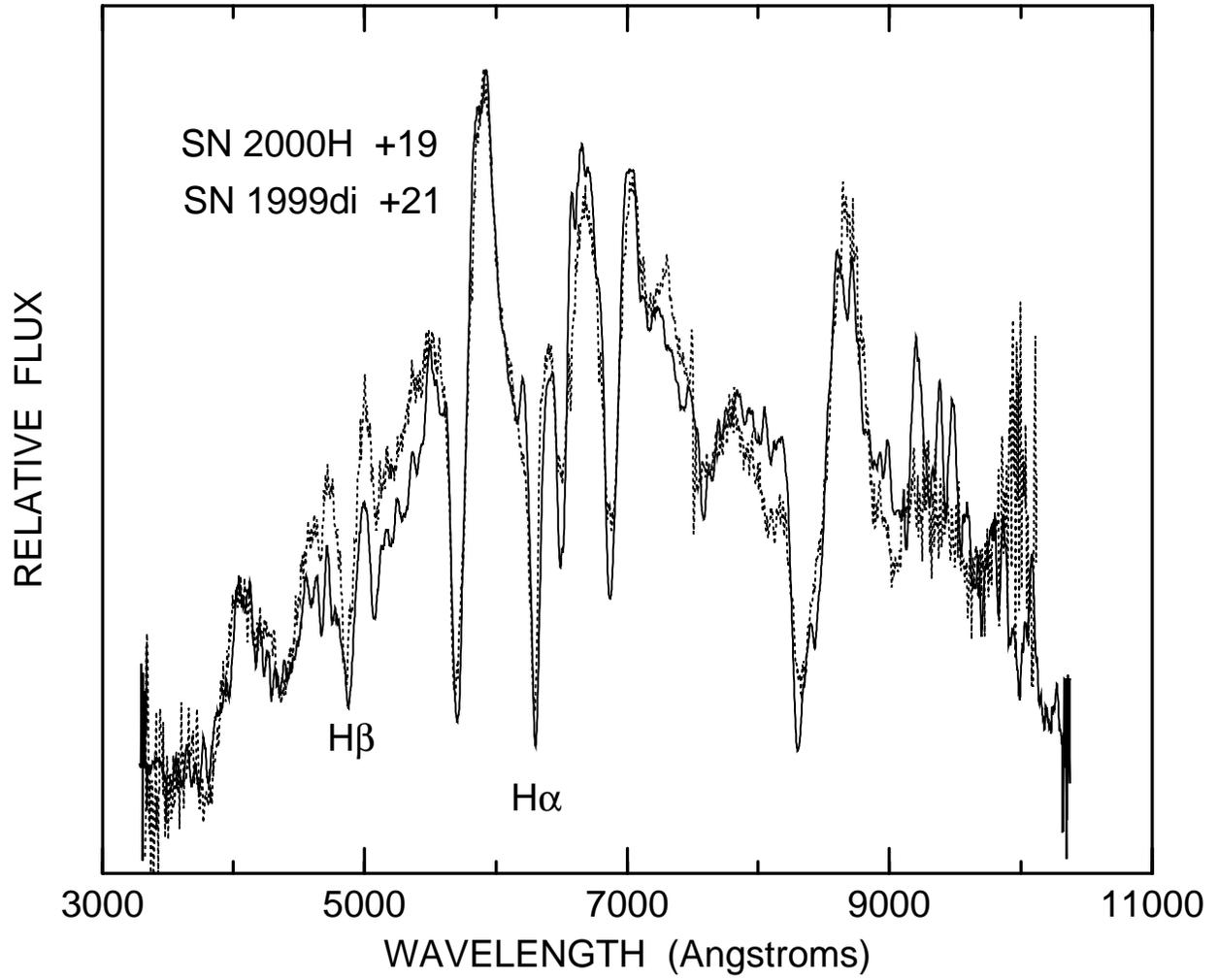}

\figcaption{The $+21$ day spectrum of SN~1999di (solid line) is compared with the
$+19$ day spectrum of SN~2000H (dotted line).}\end{figure}
\begin{figure}
\plotone{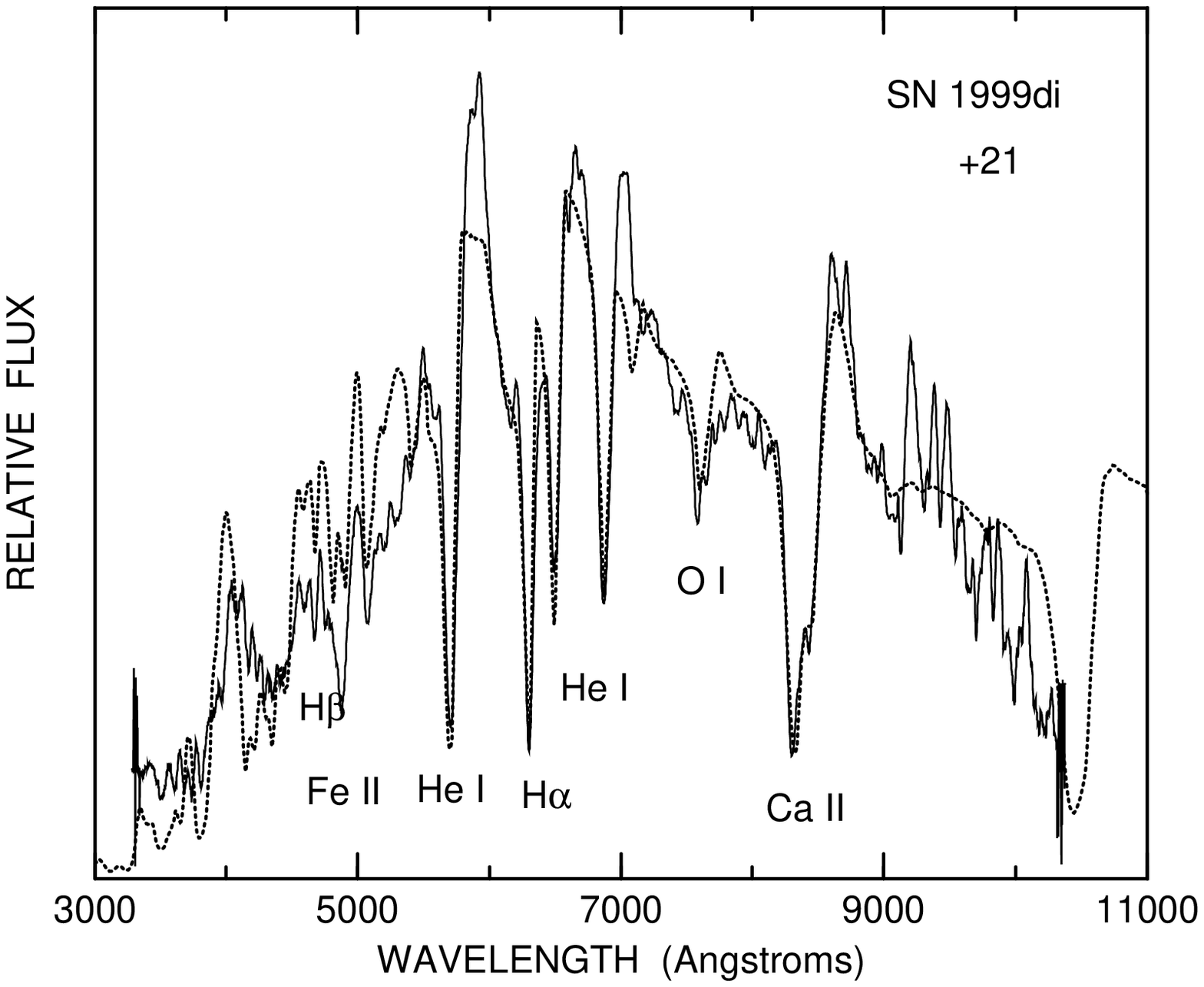}

\figcaption{The $+21$ day spectrum of SN~1999di (solid line) is
compared with a synthetic spectrum (dotted line) that has $v_{phot}=
7000$~\kms\ and $T_{bb}=4500$~K, and contains lines of H, He~I, O~I,
Ca~II, Sc~II, Ti~II, and Fe~II.}\end{figure}
\begin{figure}
\plotone{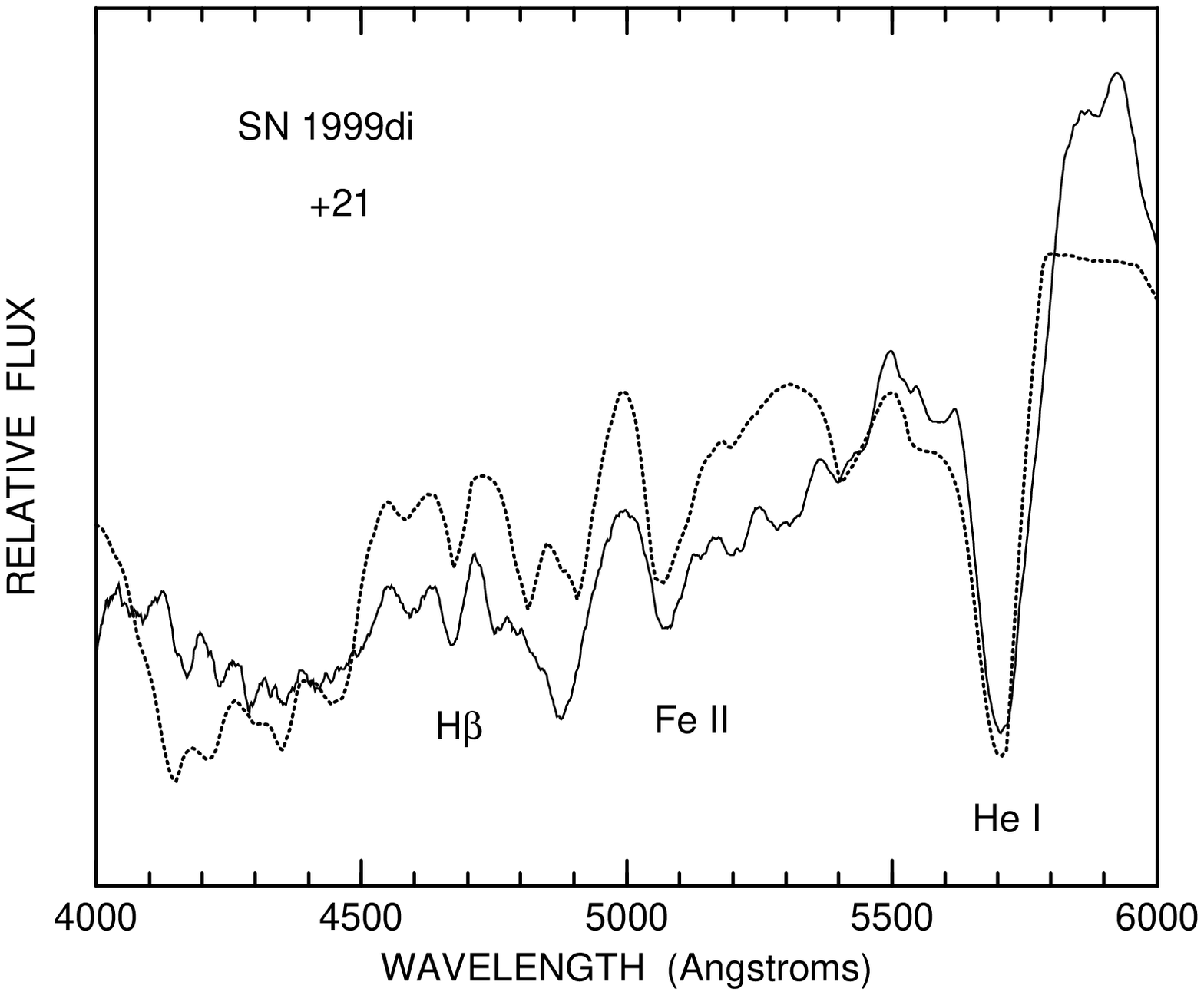}

\figcaption{Like Figure~12, but a better view of the H$\beta$ region.}\end{figure}
\begin{figure}
\plotone{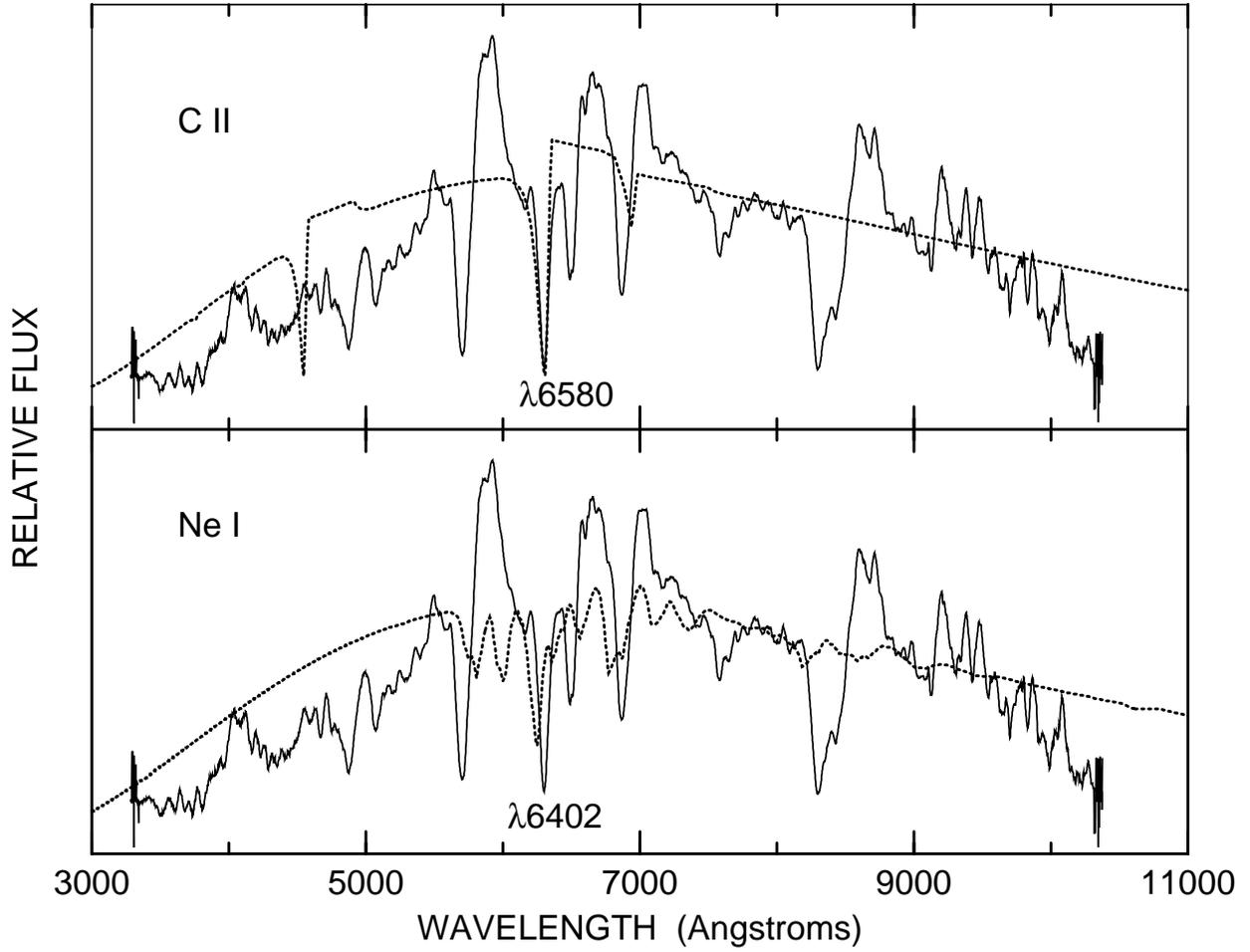}

\figcaption{The $+21$ day spectrum of SN~1999di (solid lines) is
compared with synthetic spectra (dotted lines) that contain only lines
of C~II ({\sl top panel}) and Ne~I ({\sl bottom panel}).}\end{figure}
\begin{figure}
\plotone{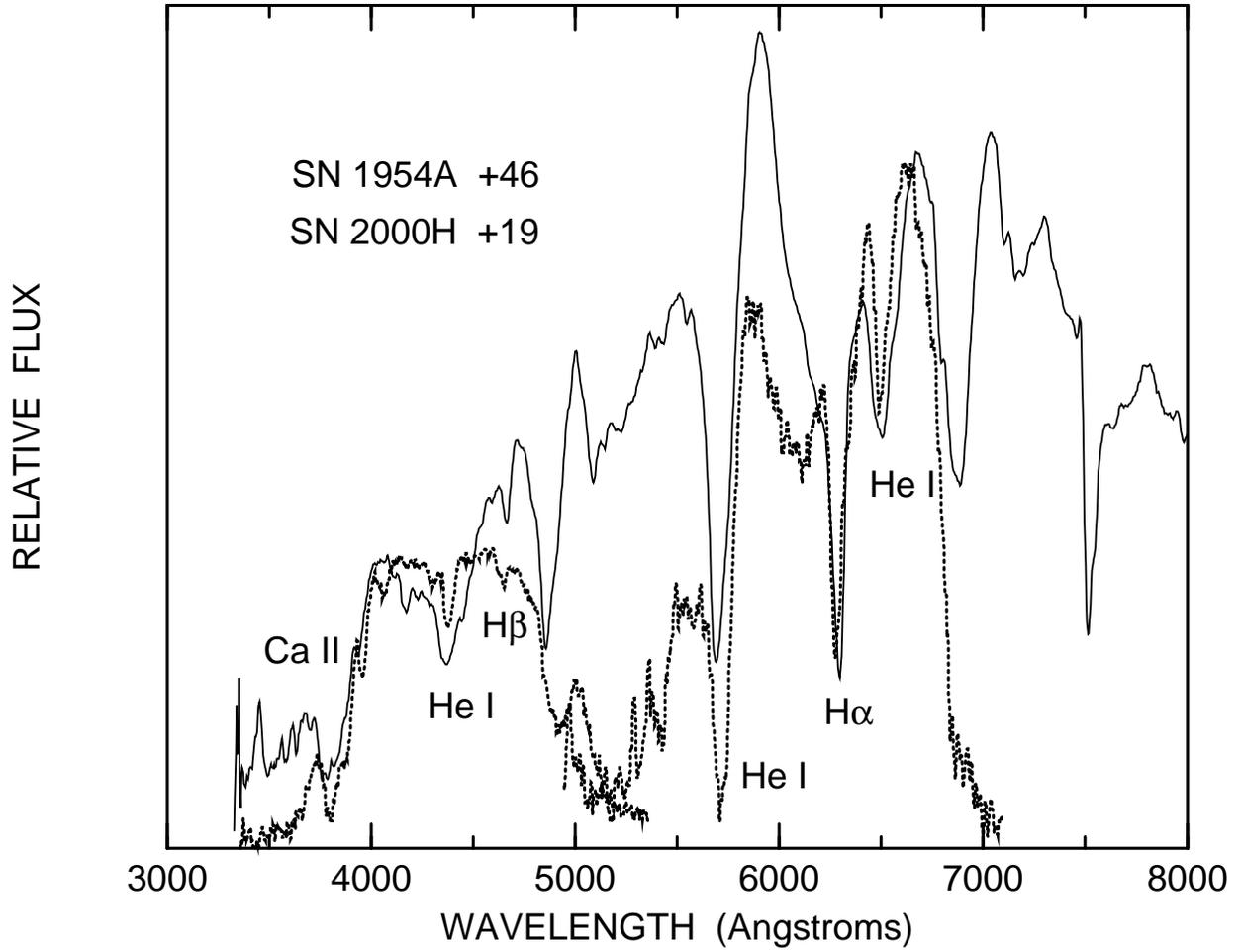}

\figcaption{Transmission tracings of two $+46$ day photographic spectra of
SN~1954A (dotted lines) are compared with the $+19$ day spectrum of
SN~2000H (solid line).}\end{figure}
\begin{figure}
\plotone{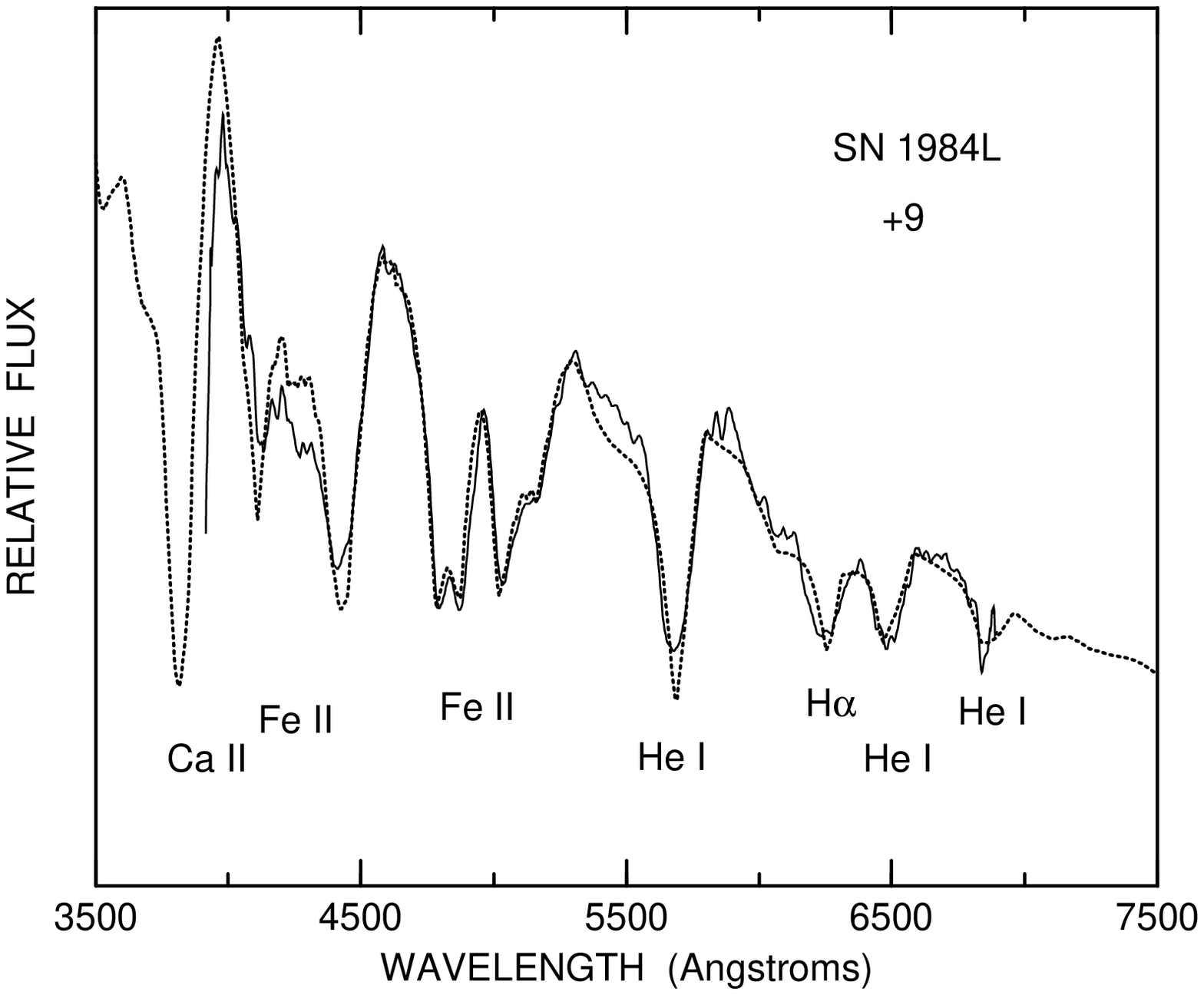}

\figcaption{The $+9$ day spectrum of SN~1984L (solid line) is
compared with a synthetic spectrum (dotted line) that has $v_{phot}=
9000$~\kms\ and $T_{bb}=8500$~K, and contains lines of H, He~I, O~I,
Ca~II, Sc~II, Ti~II, and Fe~II.}\end{figure}

\begin{figure}
\plotone{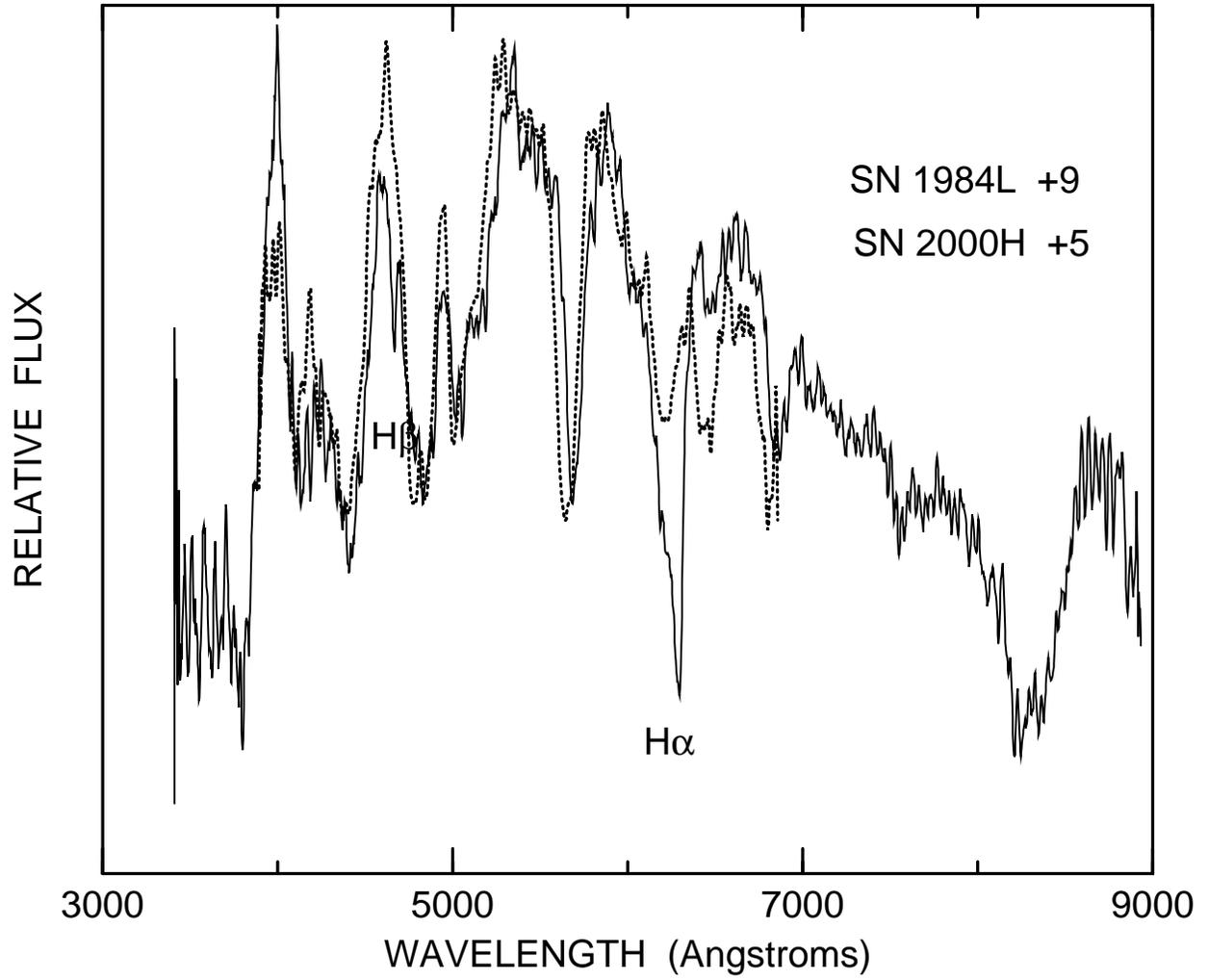}
\figcaption{The $+9$ day spectrum of SN~1984L (dotted line) 
is compared with the $+5$ day spectrum of SN~2000H (solid line).}
\end{figure}
\clearpage
\begin{figure}
\plotone{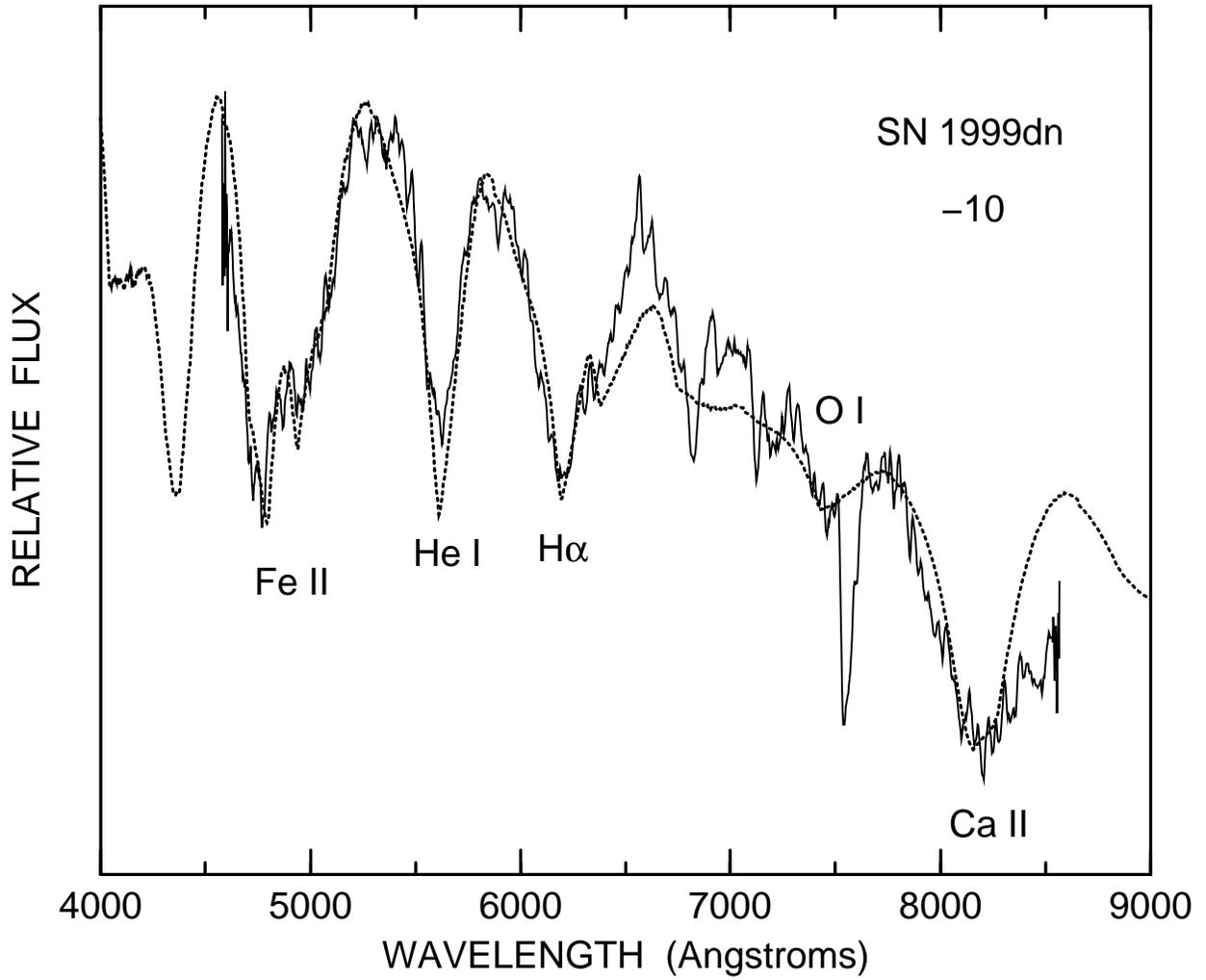}
\figcaption{The $-10$ day spectrum of SN~1999dn (solid line) is
compared with a synthetic spectrum (dotted line) that has $v_{phot}=
14,000$~\kms\ and $T_{bb}=6500$~K, and contains lines of H, He~I, O~I,
Ca~II, and Fe~II.}
\end{figure}
\clearpage
\begin{figure}
\plotone{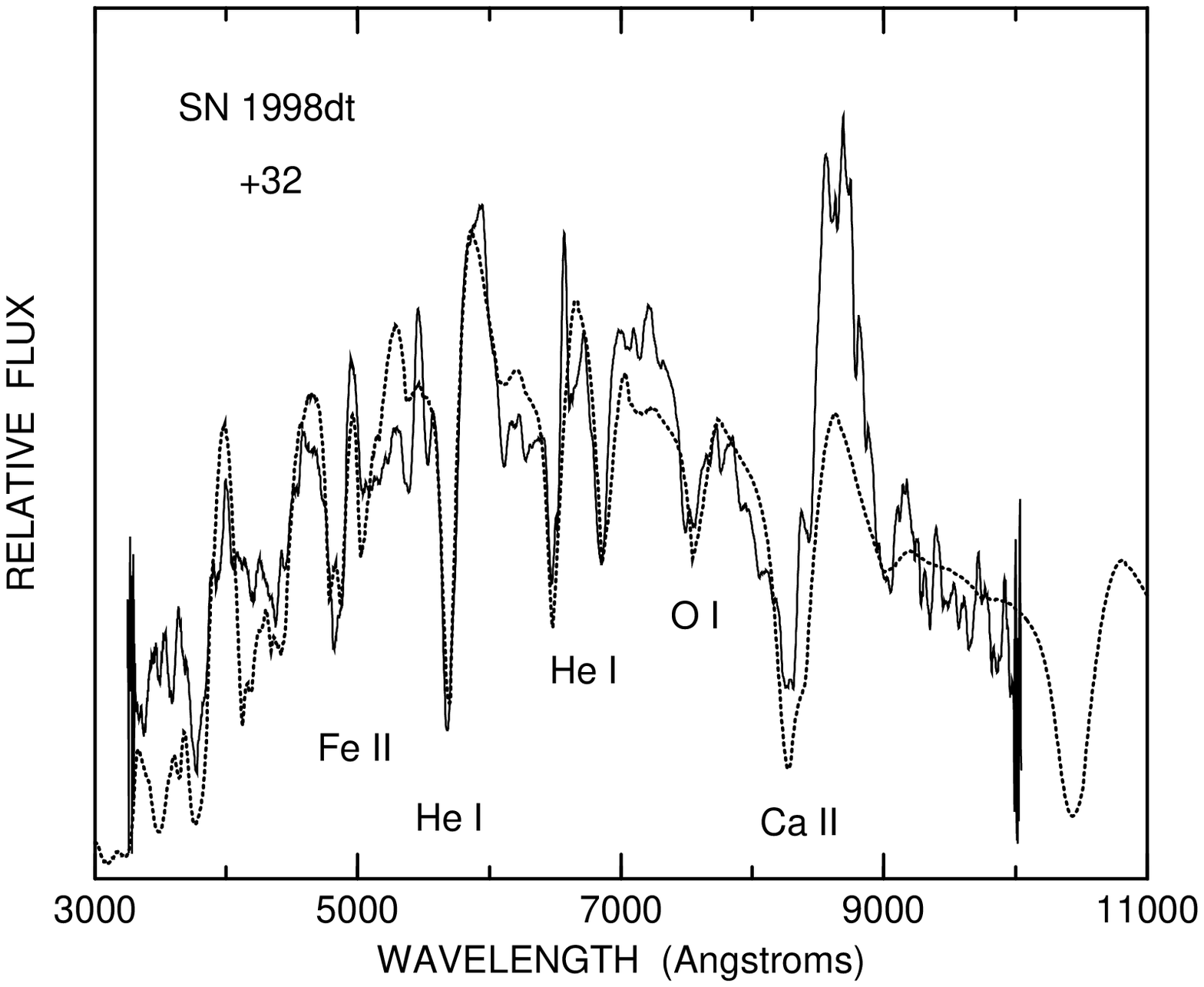}
\figcaption{The $+32$ day spectrum of SN~1998dt (solid line) is
compared with a synthetic spectrum (dotted line) that has $v_{phot}=
9000$~\kms\ and $T_{bb}=5000$~K, and contains lines of He~I, O~I,
Ca~II, Sc~II, Ti~II, and Fe~II.}

\end{figure}
\begin{figure}
\plotone{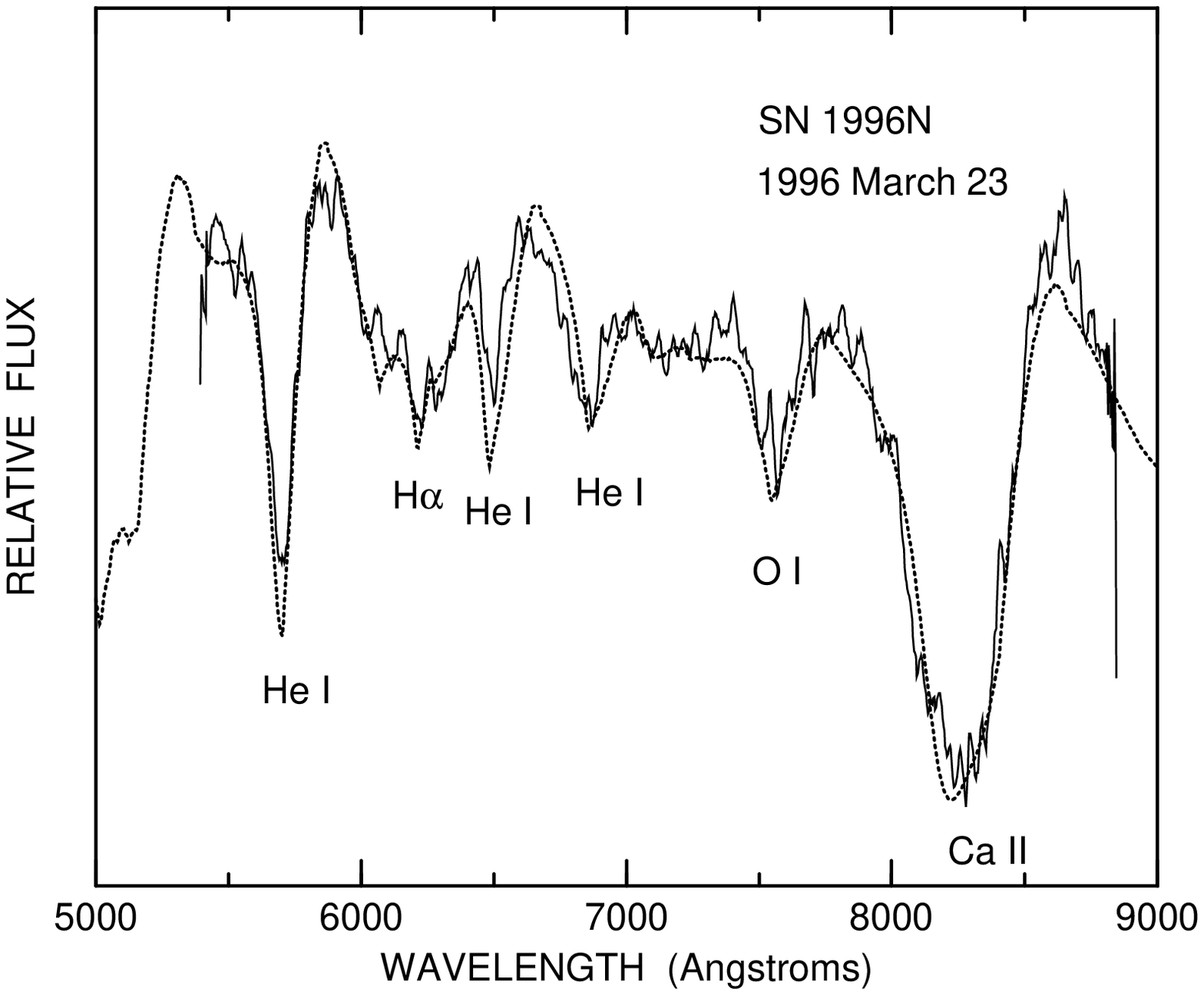}
\figcaption{The 1996 March~23  spectrum of SN~1996N (solid line) is
compared with a synthetic spectrum (dotted line) that has $v_{phot}=
9000$~\kms\ and $T_{bb}=5000$~K, and contains lines of H, He~I, O~I,
Ca~II, and Fe~II.}
\end{figure}

\begin{figure}
\plotone{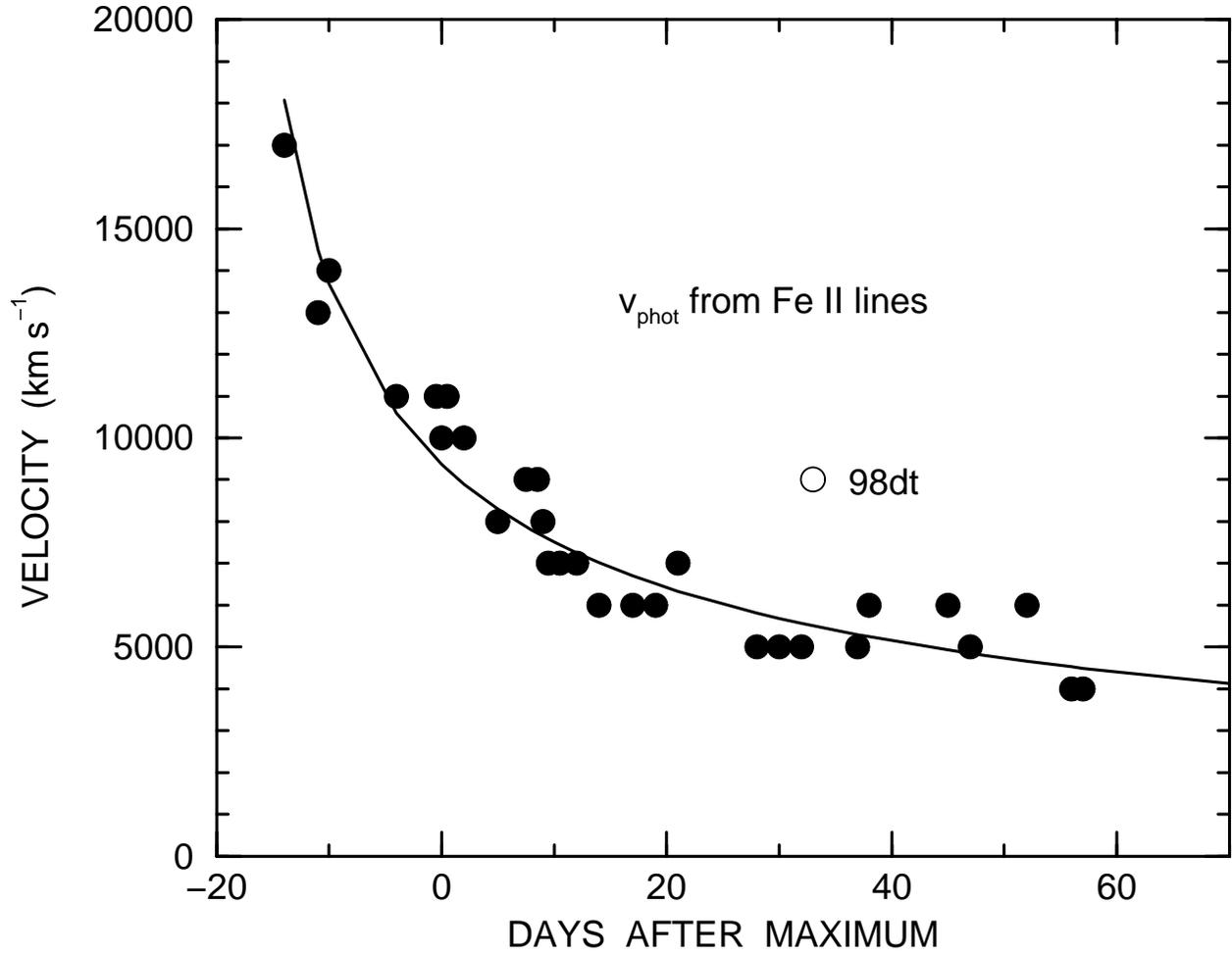}
\figcaption{The velocity at the photosphere, as inferred from Fe~II
lines, is plotted against time after maximum light.  The line is
a power--law fit to the data, with SN~1998dt at 32 days (open
circle) excluded.}
\end{figure}

\begin{figure}
\plotone{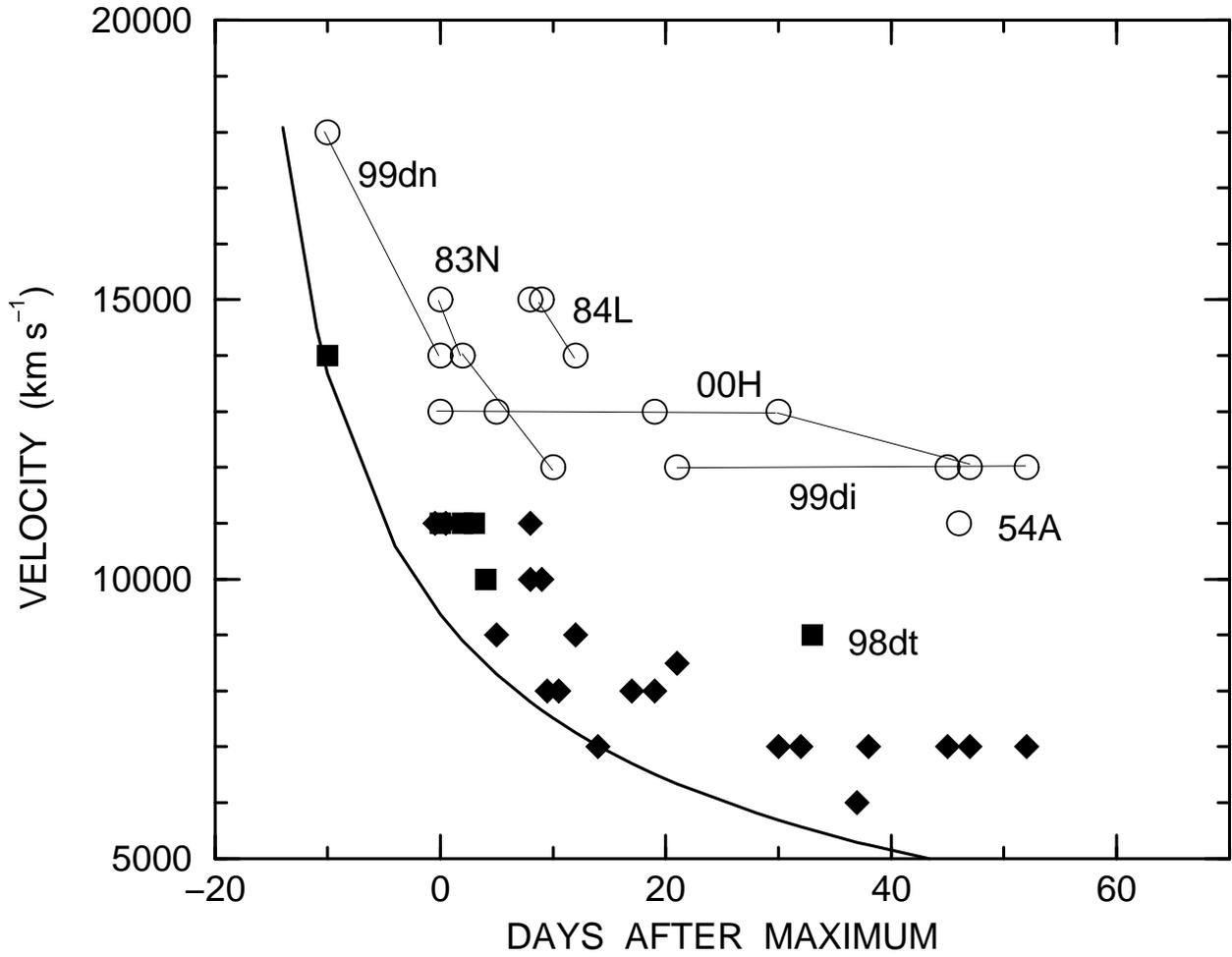}
\figcaption{The minimum velocity of the He~I lines (filled squares when
undetached, filled triangles when detached) and the minimum velocity
of the hydrogen lines (open circles; always detached) are plotted
against time after maximum light.  The curve is the power--law
fit to the velocity at the photosphere, from Figure~22.}
\end{figure}

\begin{figure}
\plotone{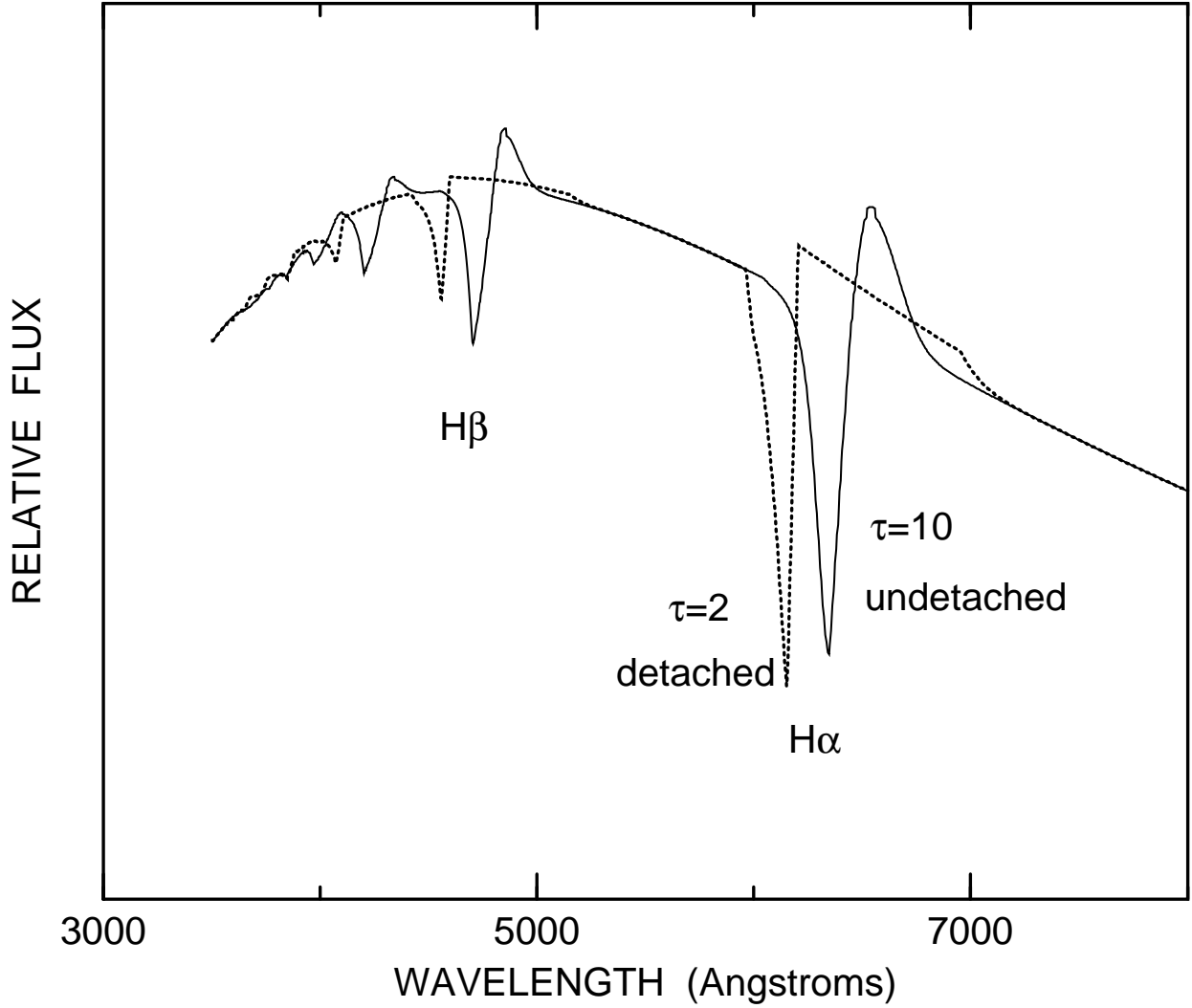}
\figcaption{A synthetic spectrum (dotted line) that has 
\vp=10,000~\kms\ and hydrogen
lines detached at 20,000~\kms, where $\tau$(H$\alpha$)=2, is compared
with a synthetic spectrum (solid line) that has
\vp=10,000~\kms\ and undetached hydrogen lines, with $\tau$(H$\alpha$)=10
at the photosphere.}
\end{figure}

\end{document}